\documentclass[10pt,oneside,onecolumn,british,useAMS,usenatbib]{aa}
\usepackage{longtable}
\usepackage{graphicx}
\usepackage[normalem]{ulem}

\makeatletter

\providecommand{\tabularnewline}{\\}

\usepackage{babel}

\usepackage[varg]{txfonts}
\makeatother

\begin{document}

\title{On the dynamical history of the interstellar
object 'Oumuamua.}

\author{Piotr A. Dybczyński\inst{\ref{inst1}} and Małgorzata Królikowska\inst{\ref{inst2}}}

\institute{Astronomical Observatory Institute, Faculty of Physics, A.Mickiewicz
University, Słoneczna 36, Poznań, Poland, \email{dybol@amu.edu.pl}\label{inst1}
\and Space Research Centre of Polish Academy of Sciences, Bartycka
18A, Warszawa, Poland, \email{mkr@cbk.waw.pl} \label{inst2}}

\date{Received xxxxx / Accepted yyyyyy}

\abstract{1I/2017~U1 'Oumuamua is the first interstellar object recorded inside
the Solar System. We try to answer the \textnormal{main} question:
where does it come from? To this aim we searched for \textnormal{close encounters} between
'Oumuamua and all nearby stars with known kinematic data during their
past motion. We had checked over 200~thousand stars and found just
a handful of candidates. If we limit our investigation to within a
60 pc sphere surrounding the Sun, then the most probable candidate
for the 'Oumuamua parent stellar habitat is the star \object{UCAC4
535-065571}. However \object{GJ 876} is also a favourable candidate.
Moreover, the origin of 'Oumuamua from a much more distant source
is still an open question. \textnormal{Additionally,} we found that the quality
of the original orbit of 'Oumuamua is accurate enough for such a study
and that none of the checked stars had perturbed its motion significantly.
All numerical results of this research are available in the Appendix.}


\titlerunning{On the dynamical history of the interstellar object 'Oumuamua}

\authorrunning{P.A.Dybczyński\&M.Królikowska}

\maketitle

\section{Introduction}\label{sec:Introduction}

\textnormal{The problem of the lack of evident interstellar visitors in our Solar System has been discussed for decades. Recently, \citet{Engelhardt_et_al:2017} considered the implications
of not observing such interstellar visitors. Now, the situation has changed.}

The first interstellar small body penetrating
our Solar System was discovered on Pan-STARRS1 images taken on Oct.\,18.5
UT at mag 19.8 (MPC CBET 4450). Initially, it was designated as a
comet (C/2017 U1) due to its near-parabolic orbit. Later on, due to
the lack of any cometary activity it was renamed as A/2017~U1 (M.P.E.C.
2017-U183, issued on 2017 Oct.~25, 22:22~UT). Ten days later, in
M.P.E.C. 2017-V17, issued on Nov.~6, 21:00~UT, a new concept for
naming such unusual objects was announced and accordingly, A/2017~U1
was renamed as 1I/2017 U1 ('Oumuamua).

The unique dynamical nature of this object was first noted by Bill
Gray in his October 25 posting to the Minor Planet Mailing list (MPML)\footnote{https://groups.yahoo.com/neo/groups/mpml/info}. He obtained
a preliminary orbit based on a six day arc and noticed an atypically
high eccentricity of approximately 1.2.\textnormal{ 'Oumuamua travels at a rather high velocity with respect
	to the Sun (on the order of 25\,km\,s$^{-1}$).} Several preprints on the kinematics
of this extraordinary object have recently appeared. \citet{Mamajek-astro-ph:2017}
analysed the stars nearest the Sun for similar spatial velocity while
\citet{Gaidos-astro-ph:2017} suggested the origin of 'Oumuamua in
a nearby young stellar cluster.

'Oumuamua seems to be unique for its physical characteristics as well.
\citet{Meech_et_al:2017} estimated its shape to be extremely elongated
while \citet{Fraser_et_al:2017} and \citet{Drahus_et_al:2017} determined
it to be a tumbling body.\textnormal{ An interesting paper on physical
parameters determination of 'Oumuamua has been also presented recently
by \citet{Jewitt_lu:2017}.}

Since the nature of this object is still unknown it might be desirable
to study its dynamical history before entering the Solar System
interior.

This paper is organized as follows: the next section describes the
model of solar vicinity dynamics which we use to track 'Oumuamua's \textnormal{past}
motion. The main task was to collect data on potential stellar perturbers.
Section \ref{sec:Results} presents the results of our numerical experiments.
In the last section, we interpret these results and discuss their
importance. \textnormal{In the Appendix we present complete numerical results, all stellar parameters used in this work with their references and several examples of the geometry of the Qumuamua - star encounters.}

\begin{figure}
\includegraphics[width=1\columnwidth]{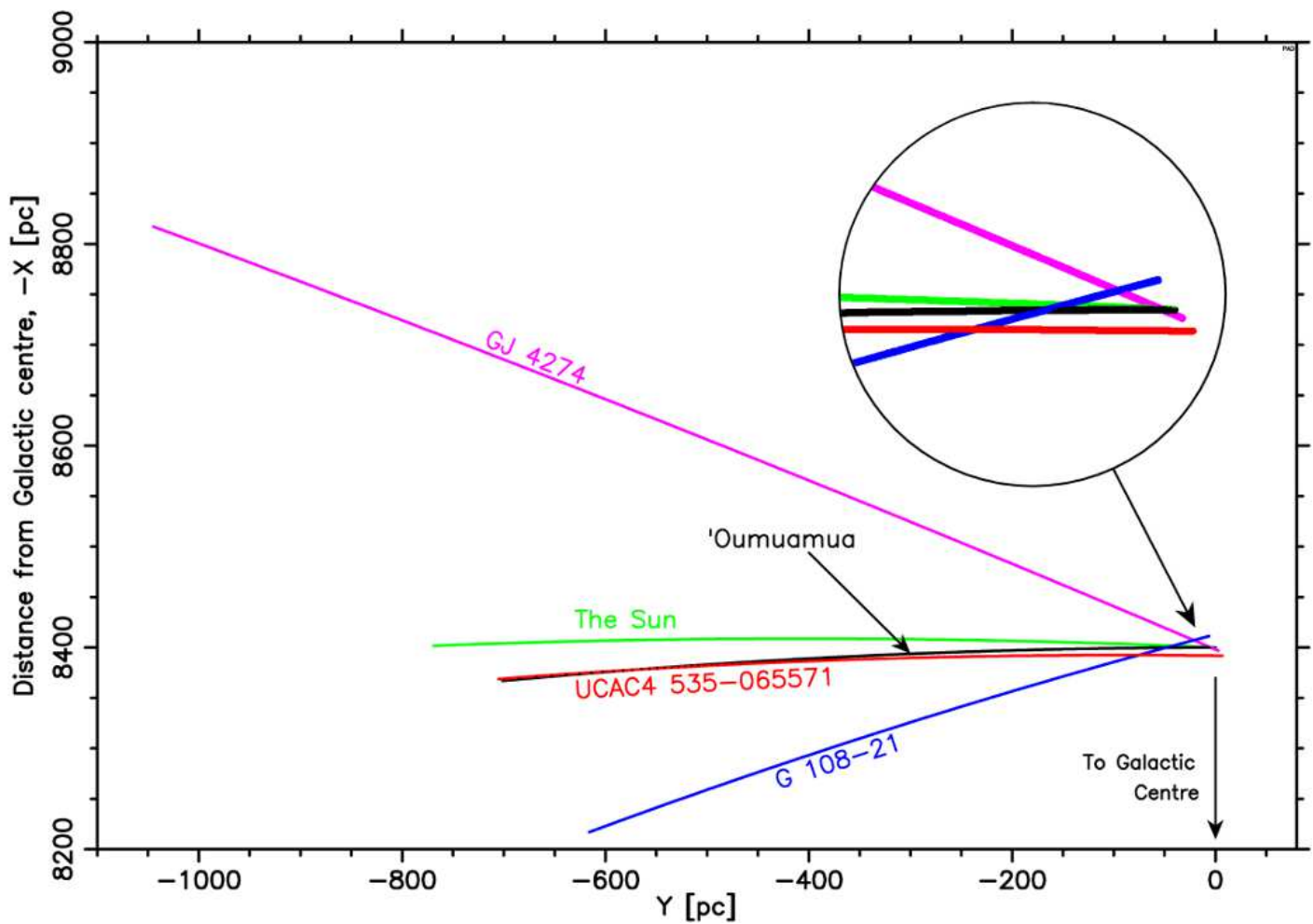}

\caption{\label{fig:three_stars}The past trajectories of 'Oumuamua, the Sun and three selected stars
during the last 3 Myr. Their positions are projected onto the XY plane
of the Galactocentric, non-rotating, right-handed rectangular frame.
This plane coincides with the Galactic Disk plane. The OX axis is
directed opposite to the Galactocentric direction to the Sun at the
starting epoch t=0.}

\end{figure}

\section{Approach to the problem}

\label{sec:Methodology}

To analyze the interstellar path of 'Oumuamua in the solar neighbourhood
it is necessary to numerically integrate its equations of motion,
taking into account both the overall Galactic potential and all important
individual stellar perturbations \textnormal{from the} known nearby stars.

To work with contemporary stellar data, we searched the whole SIMBAD
astronomical objects database\footnote{http://simbad.u-strasbg.fr/simbad/}
for all stars with known positions, proper motions, parallaxes and
radial velocities. To make sure we were working only with reliable
data, we \textnormal{demand} parallaxes to be positive and radial velocities
to be $\leqq500$ km\,s$^{-1}$ in modulus. The result of this search,
performed on November 5, 2017 \textnormal{consists of} 201\,763 individual
objects. Such a \textnormal{great} number is the result of large observational
projects, mainly Gaia \citep{Gaia_mission:2016} and RAVE \citep{RAVE-5:2017}.
\textnormal{As it concerns a homogeneity of the data taken from the SIMBAD
database we observe that 84 percent of astrometric measurements of
these 201\,763 stars were copied from TGAS catalogue \citep{2016A&A...595A...2G}
and another 11 per cent from the HIP2 catalogue \citep{vanleeuwen:2007}.
The situation is a bit more complicated with radial velocity sources
but still the majority of measurements (70 per cent) was taken from
RAVE data releases \citep{RAVE-5:2017,RAVE-4:2014} and another 11
per cent from the Pulkovo compilation \citep{Gontcharov:2006}. The
remaining radial velocity measurements taken by us from the SIMBAD
database were copied from a large number of individual papers.}

This stellar data set allowed us to perform accurate calculations,
namely the numerical integration of 'Oumuamua's motion. To account
for mutual stellar gravitational interactions, we had to integrate
the N-body problem, consisting of 'Oumuamua, the Sun and all individual
stellar perturbers, all of which are under the influence of the overall
Galactic potential. However, integrating the simultaneous motion's
of over 200 thousand bodies would be a waste of time - a great majority
of these bodies never came close enough to 'Oumuamua to disturb its
motion. Instead we first prepared a short list of perturbers, see
below.

\textnormal{Numerical integration of motion was performed in a right-handed,
non-rotating, rectangular Galactocentric frame with the OX axis directed
in the opposite direction to the Sun's position at the starting time.
We use the Model I variant of the Galactic gravitational potential
described by \citet{irrgang_et_al:2013}. As the starting position
and velocity of the Sun we use vectors: $\mathbf{R_{\odot}}=(x_{\odot},y_{\odot},z_{\odot})=(-8400,0,17)$
in pc and $\mathbf{\dot{R}}_{\odot}=(u,v,w)=$ (+11.352, +260.011,
+7.41) in pc/Myr. In the former we adopted the vertical position of
the Sun given by \citet{joshi:2007}. }A detailed description of the
Galactic reference frame orientation \textnormal{and }the Galactic potential
form and parameters can be found in \citet{dyb-berski:2015}. \textnormal{Here
we use }exactly the same dynamical model and equations of motion.

The starting position and velocities of 'Oumuamua for dynamical calculations
outside \textnormal{a} planetary zone were obtained from its original
orbit. We determined it from all positional observations available
in the MPC database \footnote{http://www.minorplanetcenter.net/db\_search}
on November 12, 2017. Through careful data processing, we obtained
an osculating orbit given in Table~\ref{tab:elem_oskul}. Next, in
order to observe the uncertainties in the motion of 'Oumuamua at every
stage of our research, we cloned its orbit and built a swarm of 10\,000
orbits resembling the observations, using a method described by \citet{sitarski:1998}
\textnormal{which fully utilises the covariance matrix obtained during
the orbit determination.} Then, we numerically propagated all these
orbits forward and backward up to a heliocentric distance of 250\,AU
(the distance at which planetary perturbations are negligible).
The resulting barycentric elements of the original and future orbits
of 'Oumuamua, along with their uncertainties, are presented in Table
\ref{tab:elem_ori_fut}. The barycentric positions and velocities
of each individual clone of 'Oumuamua at 250 au were used by us as
starting data for a \textnormal{dynamical} study of this body \textnormal{under}
the gravitational\textnormal{ influence }of stars and the full Galactic
potential.

We suppose that due to the lack of cometary activity, nongravitational
forces (\textnormal{we found them} non-detectable from positional data)
could not have changed the orbit of 'Oumuamua significantly during
its close perihelion \textnormal{passage} and that the original orbit
is rather reliable, with the uncertainties presented in Table~\ref{tab:elem_ori_fut}.
To observe how \textnormal{these} uncertainties influence the
minimal distance between 'Oumuamua and all stars included in our model,
we repeated our numerical integration for all 10\,000 clones of 'Oumuamua.
\textnormal{Each encounter} parameters obtained from this complex calculation\textnormal{
as well as their variation intervals} are presented in Table \ref{A3}
in the Appendix.

\begin{table*}
	\caption{\label{tab:elem_oskul}Osculating heliocentric orbit of 'Oumuamua,
		based on 118 positional observations spanning the interval from 2017–10–14
		to 2017–11–10, available at MPC on November 12, 2017. Equator and
		ecliptic of J2000 is used. The obtained RMS is 0.35 arcsec.}
	
	\centering{}%
	\begin{tabular}{lr}
		\hline 
		perihelion distance {[}AU{]}  & 0.255234 $\pm$ 0.000062\tabularnewline
		eccentricity  & 1.199236 $\pm$ 0.000164\tabularnewline
		inverse of the semimajor axis {[}AU$^{-1}${]}  & -0.780603 $\pm$ 0.000618\tabularnewline
		time of perihelion passage {[}TT{]}  & 2017–09–09.488519 $\pm$ 0.001243\tabularnewline
		inclination {[}deg{]}  & 122.677069 $\pm$ 0.005823\tabularnewline
		argument of perihelion {[}deg{]}  & 241.683487 $\pm$ 0.011254\tabularnewline
		longitude of the ascending node {[}deg{]}  & 24.599729 $\pm$ 0.000264\tabularnewline
		epoch of osculation {[}TT{]}  & 2017 Sep. 4.0 TT = JD 2458000.5\tabularnewline
		\hline 
	\end{tabular}
\end{table*}

However, the most important source of the \textnormal{close passage} distance uncertainty,
not estimated in this paper, is \textnormal{the} stellar data errors.
This cannot be simply modelled by the simultaneous drawing of $N$
clones for all 57 stars and 'Oumuamua because that would require $N^{58}$
numerical integrations. In this paper, we restrict the error budget
calculations to the influence of the 'Oumuamua orbit uncertainty.

To refine (and considerably shrink) our set of stellar perturbers,
we first numerically followed the \textnormal{past} motion of 'Oumuamua with each of
the 201\,763 stars individually along with the Sun, forming a 3-body
problem under the influence of the full Galactic potential. During
this preliminary calculation, we assumed all stellar masses to be
1.0 M$_{\odot}$. Using these results, we selected 109 stellar objects
that passed 'Oumuamua closer than 3.5\,pc. The parameters of all
these encounters, derived from a nominal 'Oumuamua orbit and nominal
stellar data, can be found in Table \ref{A1} in the Appendix.

After a detailed inspection, involving the removal of obsolete objects
and replacing components of multiple stellar systems with their respective
centre of mass parameters, we finally collected a list of 57~stars
or stellar systems which should be taken into account when studying
'Oumuamua's \textnormal{past} motion in the Solar neighbourhood. To use
these stars as perturbers it was necessary \textnormal{to find estimations
of their} masses. \textnormal{It appeared that a} lot of them are red
(or even brown) dwarfs with \textnormal{a} very small \textnormal{mass}. Additionally, we recognised
several pairs of stars forming double systems as well as one triple
system (Alpha Centauri A,B + \object{Proxima}) and calculated their
centre of mass \textnormal{coordinates}, total mass and a systemic velocity.
The most massive perturbers in our list are the \object{Alpha Centauri}
and \object{Sirius} systems. A list of these perturbers \textnormal{with
their estimated masses, starting positions and velocities} is presented in Table \ref{A2}
in the Appendix.\textnormal{ In the last column of this table we present
references for all values used by us.} Some adopted mass values are
rather crude estimations, but due to 'Oumuamua's large velocity it
turned out that the change in a perturbers mass does not significantly
influence the path of 'Oumuamua. This of course might be false for
perturbers mutual interactions.

Finally, we integrated the N-body problem, consisting of 'Oumuamua,
the Sun and all 57 individual stellar perturbers, a 59-body system
under the influence of the overall Galactic potential (hereafter 59B
model).

\textnormal{Figure~\ref{fig:three_stars} shows the past trajectories of 'Oumuamua (in black), the Sun (in green) and three example stars selected from Tables~\ref{tab:proximities}--\ref{tab:low_vel}. Their motion is projected onto the Galactic Disk plane.
}

\begin{table*}
\caption{\label{tab:elem_ori_fut}Barycentric original and future 'Oumuamua
orbit elements.}

\centering{}%
\begin{tabular}{lrr}
\hline 
\multicolumn{1}{c}{element}  & \multicolumn{1}{c}{original orbit}  & \multicolumn{1}{c}{future orbit} \tabularnewline
\hline 
\hline 
perihelion distance {[}AU{]}  & 0.252062 $\pm$ 0.000063  & 0.257286 $\pm$ 0.000063\tabularnewline
eccentricity  & 1.196488 $\pm$ 0. 000164  & 1.200366 $\pm$ 0.000167\tabularnewline
time of perihelion passage {[}TT{]}  & 2017–09–09.118037 $\pm$0.001262  & 2017–09–09.310111 $\pm$ 0.001250\tabularnewline
inverse of the semimajor axis {[}AU$^{-1}${]}  & -0.779521 $\pm$ 0.000456  & -0.778771 $\pm$ 0.000455\tabularnewline
inclination {[}deg{]}  & 122.725937 $\pm$ 0.005995  & 122.870243 $\pm$ 0.005900\tabularnewline
argument of perihelion {[}deg{]}  & 241.696866 $\pm$ 0.011361  & 241.842028 $\pm$ 0.011433\tabularnewline
longitude of the ascending node {[}deg{]}  & 24.251515 $\pm$ 0.000251  & 24.747600 $\pm$ 0.000256\tabularnewline
epoch of osculation {[}TT{]}  & 1973–10–05  & 2061–08–04\tabularnewline
\hline 
\end{tabular}
\end{table*}

\section{Results\label{sec:Results}}

\begin{figure*}[t]
\includegraphics[angle=270,width=1\textwidth]{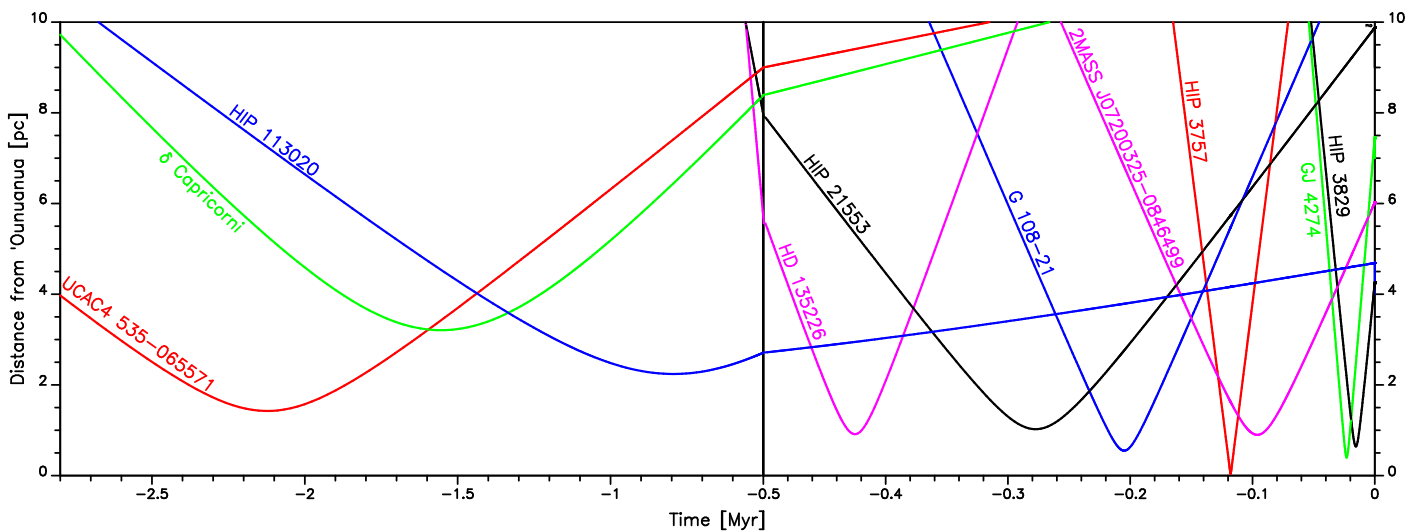}

\caption{\label{fig:startrek}Changes in the distance between 'Oumuamua and stars listed in Tables
\ref{tab:proximities} and \ref{tab:low_vel}. Omitted are only HIP~981
and TYC~5325-1808-1 due to their unreliable kinematic data. Please
note a horizontal scale change in the middle of the plot.}
\end{figure*}

After analysing 'Oumuamua's \textnormal{past} motion within the Solar
vicinity, we found seven\textnormal{ encounters closer} than 1~pc between
'Oumuamua and nearby stars. These encounters are described in Table
\ref{tab:proximities}\textnormal{ and presented in Fig.~\ref{fig:startrek}.} The first 'Oumuamua – star encounter \textnormal{(with HIP\,3757)} is a
very close one, with a miss distance of only 0.04 pc but with a rather
high relative velocity of over 200\,km s$^{-1}$. It happened 118
thousand years ago. The second encounter, with \object{GJ 4274},
happened only 23\,000 years ago with an even greater relative velocity
of 316\,km s$^{-1}$. The \textnormal{minimum distance} of the third
\textnormal{event}, with \object{HIP 981}, is also very close, but
due to the large heliocentric distance of this event and practically
unknown radial velocity of the star (rv=4.00$\pm$6.5\,km s$^{-1}$,
\citep{barbier-brossat-f:2000}) we treat this case as a ``false
positive''. The remaining cases presented in Table \ref{tab:proximities}
yield a relative velocity of over 60 km s$^{-1}$, which also makes
them not very promising candidates for 'Oumuamua's source system.

\begin{table*}
\caption{\label{tab:proximities}All \textnormal{close encounters} between 'Oumuamua and a star
or stellar system closer than 1\,pc obtained from the 59B model. Only
the results for \object{HIP 981} are from a 3-body calculation,
see text for the explanation. \textnormal{The star HIP~21553 is additionally
added since its min distance is only slightly over 1~pc and it has
a small relative velocity.} The nominal orbit of 'Oumuamua and nominal
stellar data are used for the results presented in columns 2–4. In
the last column we show the minimum proximity distance variation interval
obtained from the 10\,000 clones of 'Oumuamua.}

\centering{}%
\begin{tabular}{lrrrrc}
\hline 
Star name  & min distance  & epoch  & rel velocity  & r\_hel  & miss-distance interval\tabularnewline
 & {[}pc{]}  & {[}Myr{]}  & {[}km s$^{-1}${]}  & {[}pc{]}  & {[}pc{]}\tabularnewline
\hline 
\object{HIP 3757}  & 0.04401  & -0.1179  & 185.375  & 3.175  & 0.04192 – 0.04670\tabularnewline
\object{GJ 4274}  & 0.41190  & -0.0227  & 309.498  & 0.612  & 0.41123 – 0.41257\tabularnewline
\object{HIP 981}  & 0.50883  & -6.6740  & 17.660  & 178.558  & not tested\tabularnewline
\object{G 108-21}  & 0.55083  & -0.2046  & 64.602  & 5.507  & 0.54261 – 0.55933\tabularnewline
\object{HIP 3829}  & 0.64437  & -0.0152  & 240.683  & 0.411  & 0.64422 – 0.64460\tabularnewline
\object{2MASS J07200325-0846499}  & 0.90002  & -0.0953  & 62.902  & 2.567  & 0.89864 – 0.90144\tabularnewline
\object{HD 135226}  & 0.92358  & -0.4269  & 68.669  & 11.491  & 0.92010 – 0.92795\tabularnewline
\object{HIP 21553}  & 1.01326 & -0.2772 & 34.723 & 7.453 & 1.01326 – 1.03281\tabularnewline
\hline 
\end{tabular}
\end{table*}

\begin{table*}
\caption{\label{tab:low_vel}Four cases of low velocity encounters of 'Oumuamua
with stars from our list. Columns 2–4 are from the 59B calculation.
Only the results for \object{TYC 5325-1808-1} are from the 3-body
integration, see text for the explanation. The last column presents
the miss-distance variation interval from the 10\,000 clones of 'Oumuamua.
Distances are in parsecs, times in Myr and velocities in km s$^{-1}$.}

\centering{}%
\begin{tabular}{lrrrrc}
\hline 
Star name  & miss-distance  & epoch  & rel velocity  & r\_hel  & miss-distance interval\tabularnewline
 & {[}pc{]}  & {[}Myr{]}  & {[}km s$^{-1}${]}  & {[}pc{]}  & {[}pc{]}\tabularnewline
\hline 
TYC 5325-1808-1  & 2.93291  & -10.1215  & 7.940  & 269.220  & not tested\tabularnewline
UCAC4 535-065571  & 1.42718  & -2.1395  & 5.206  & 57.538  & 1.37201 – 1.48409\tabularnewline
$\delta$Capricorni  & 3.21225  & -1.5259  & 6.872  & 41.049  & 3.16832 – 3.25921\tabularnewline
HIP 113020  & 2.24104  & -0.7907  & 3.927  & 21.278  & 2.21875 – 2.26360\tabularnewline
\hline 
\end{tabular}
\end{table*}

When searching for the parent star of 'Oumuamua, one probably should
look for a close passage with a much smaller relative velocity. From
among the over 200\,000 tested stars, we found only four such cases,
see Table~\ref{tab:low_vel}.

\textnormal{In Fig.~\ref{fig:startrek} we show how the distance of 'Oumuamua from stars listed in Table~\ref{tab:proximities} and Table~\ref{tab:low_vel} changed in time. }

Almost 820 thousand years ago, 'Oumuamua passed near the star \object{HIP
113020} (also known as \object{BD-15 6290}, \object{GJ 876},
or \object{Ross 780}) with a relative velocity of about 5\,km
s$^{-1}$ and at a heliocentric distance of 21.3~pc. For the nominal
'Oumuamua orbit, the minimal distance between these two objects was
2.24 pc. However, it should be noted that (going backwards along its
track) the motion of 'Oumuamua was perturbed by six stars from Table
\ref{tab:proximities} as well as 51 other stars acting from larger
distances of 1 – 3 pc. Every close passage of 'Oumuamua near a star
magnifies starting point uncertainties,\textnormal{ additionally increased
by the stellar data errors}. While most of the stars included in our
calculations are M dwarfs with rather small masses, some have masses
greater than the Sun. However, while the stellar kinematics data uncertainties
are the most important source of the proximity distance uncertainty,
these are not estimated in this paper. To observe how the uncertainties
of 'Oumuamua's orbit affect our results, we repeated our calculation
of the 59B model for the 10\,000 clones of 'Oumuamua. We individually
searched for the closest and the farthest clone at the approach epoch
and recorded the \textnormal{encounter} parameters for each star, obtaining
their variation intervals. For this number of clones, these intervals
are wider than 3$\sigma$ and are presented in the last column of
Table \ref{tab:proximities}. Similar data for all the studied stars
may be found in Table \ref{A3} in the Appendix.

Three more low velocity encounters happened further than 30~pc from
the Sun. We recognised the encounters with the high proper motion
star UCAC4~535-065571 and the eclipsing binary $\delta$Capricorni
as the most interesting ones. \object{TYC 5325-1808-1} cannot be
reliably included in our list of perturbers since its mass and spectral
type are unknown. The correct mass value is indispensable for dynamically
tracing such a long trip (almost 270~pc).

\object{UCAC4~535-065571} is a red dwarf of M6V spectral type
and its mass is estimated to be 0.205\,M$_{\odot}$\citep{Newton-et-al:2016}
We obtained\textnormal{ an encounter} relative velocity of 5.35\,km s$^{-1}$
but with a\textnormal{ minimum} distance of 3.46\,pc. With such a large
miss-distance, one might reject this star as a parental candidate
for 'Oumuamua. However, we noticed that the kinematics of \object{UCAC4~535-065571}
is rather poorly known. In the SIMBAD database we found its: parallax
plx=85.40$\pm$3.30 mas\citep{2014ApJ...784..156D}, proper motions:
pma=$-$107.0$\pm$8 mas/yr and pmd=$-$133.0$\pm$8 mas/yr \citep{2012yCat.1322....0Z}
and radial velocity rv=$-$19$\pm$5\,km s$^{-1}$\citet{2014AJ....147...20N}.
By manipulating numbers within their uncertainties, we obtained a
miss-distance of 0.6\,pc but with a relative velocity of 10\,km
s$^{-1}$, \textnormal{(by adopting:} plx=82.1 mas, pma=$-$99\,mas/yr,
pmd=141\,mas/yr and rv=14.0~km s$^{-1}$). Alternative kinematic
parameters for this star can \textnormal{also} be found in \citet{West-et-al:2015},
where plx=76\,mas and rv=$-$9.5\,km s$^{-1}$. The discrepancy
between radial velocity measurements might be connected with the rotational
velocity of 43\,km s$^{-1}$ \citep{Newton-et-al:2016} for this
star. Using these kinematic data, we obtained a nominal proximity
distance of 0.4\,pc but with a relative velocity of 14.7\,km s$^{-1}$.

$\delta$Capricorni (HIP~107556, GJ~837) is also a good candidate,
with high precision kinematic parameters. It is an eclipsing binary
so its mass is also accurate \textnormal{and it has} a small relative
velocity of 6.9 km s$^{-1}$. 'Oumuamua passed this star at a rather
large distance of 3.21 pc.

\section{Discussion and conclusions \label{sec:Discussion}}

No obvious parent star has been identified.

The closest 'Oumuamua – star proximity found by us, an encounter with
\object{HIP 3757} almost 120 thousand years ago, does not indicate
that 'Oumuamua originated from this star system. It might be true
provided some mechanism of ejecting 'Oumuamua from this system with
the relative velocity of 185\,km s$^{-1}$ would be proposed.

It seems more reasonable to search for the parent star of 'Oumuamua
in the cases \textnormal{of} a much smaller relative velocity. Utilizing
such an approach would make \object{HIP 113020} a more promising
candidate. This well known star (known also as BD-15 6290, GJ~876,
or Ross~780) has a rich planetary system \textnormal{consisting of four
planets: one very small and three other massive and strongly interacting
planets,} see for example \citet{Rivera-et-al:2010} and the references
therein. Our resulting \textnormal{miss-distance} is highly sensitive
to the systemic radial velocity of the \object{HIP 113020} system.
There seems to be some discrepancy between the centre of mass velocity
of about 0.5\,km s$^{-1}$ obtained in the paper quoted above and
the value presented in the SIMBAD database: $-$1.519$\pm$0.157\,km
s$^{-1}$\citep{2015ApJ...802L..10T}. Taking into account that the
motion of 'Oumuamua was perturbed by tens of stars after passing HIP\,113020
and that kinematic parameters (and masses) of these perturbers are
of a significantly different quality and accuracy, we cannot rule
out the possibility that 'Oumuamua originated from the \object{HIP
113020} planetary system. Definitively, we have to wait for much
more precise stellar data from the Gaia mission \citep{Gaia_mission:2016}.

In our results obtained for the vicinity of the Sun there is yet another
nearby star worth mentioning. 'Oumuamua nominally passed \object{HIP
21553} at a distance of 1.02287\,pc almost 280 thousand years ago
with a relative velocity of less than 35\,km s$^{-1}$. \object{HIP
21553} (also known as \object{HD 232979} or \object{GJ 172})
is a M0.5V type red dwarf according to SIMBAD \citep{1989ApJS...71..245K}. Its astrometric data were recently highly improved by the Gaia mission
\citep{Gaia_mission:2016}.

Additionally, we found another candidate past the 30 pc heliocentric
distance, the star UCAC4~535-065571. By varying this star's position
and velocity within their respective uncertainty intervals, we obtained
a very close encounter with 'Oumuamua at a reasonably small relative
velocity of 5–15 km s$^{-1}$. It seems necessary to study the kinematics
of this star in more detail in order to make any definitive conclusion
on the putative relation between this star and 'Oumuamua. There is
also a small probability that 'Oumuamua comes from \object{$\delta$Capricorni}.

An equally interesting hypothesis is that this interstellar object
came to our planetary system from a more distant source.

\textnormal{During the studied a few million years to the past, the heliocentric trajectory of ‘Oumuamua appeared to be almost the straight line with an approximately  constant velocity.
This is mainly because of its great velocity, relatively
large distances to perturbing stars and their small masses in the most of cases. This fact is illustrated in several figures included in Table~\ref{geometries} in the Appendix. One can also see in Fig.~\ref{fig:three_stars} that the deviation from the straight line motion is also very slight in the Galactocentric frame over the similar time interval. }

Another important conclusion from our work is that
despite the short observational interval of 'Oumuamua, its original
orbit uncertainties do not influence our results in any significant
way.

There recently appeared a preprint by S. Portegies Zwart and colleagues
\citep{Portegies-Zwart-et-al:2017} in which the authors propose five
other stars as potential sources for 'Oumuamua. We carefully checked
all these cases and according to our calculations 'Oumuamua did not
come closer than 20\,pc to any of these stars. A probable reason for
this is that an approximate dynamical model was used in the quoted
paper, where as in all five cases these stars are very distant and
therefore their motion is sensitive to the dynamical model details,
especially the mutual interactions between all stars involved. Such
a disagreement is also noted in \citet{Feng-Jones:2017}. On the contrary,
these authors confirm our results for overlapping stars. 
\begin{acknowledgements}
This research was partially supported by the project 2015/17/B/ST9/01790
founded by the National Science Centre in Poland. This research made
use of the SIMBAD database, operated at CDS, Strasbourg, France 
\end{acknowledgements}

\bibliographystyle{aa}
\bibliography{moja23,data_sources}

\newpage{}{\onecolumn{

\appendix

\section{Supplementary material}

\begin{longtable}{llrrrr}
\caption{\label{A1} Here we present 109 individual objects selected form the
SIMBAD data that fulfil the following conditions: an object must
approach 'Oumuamua closer than 3.5 pc before leaving the heliocentric
sphere of 300~pc radius. Stars are sorted here by the proximity distance.
To speed up the selection procedure the motion of each object was
traced individually as a 3-body problem under the influence of the
full Galactic potential. For its nominal orbit, we recorded the minimum
'Oumuamua – star distance (Min dist), moment of this approach (Epoch),
relative velocity (Vrel), and heliocentric distance of the approach
(r\_hel). A proximity epoch equalling zero means that the star's closest
position was at the beginning of the backward numerical integration
when 'Oumuamua was at 250 AU from the Sun.}
\tabularnewline
\hline 
Star Name  & Alternative name  & Min dist  & Epoch  & Vrel  & r\_hel\tabularnewline
 &  & {[}pc{]}  & {[}Myr{]}  & {[}km s$^{-1}${]}  & {[}pc{]} \tabularnewline
\endfirsthead

\caption{continued.}
\tabularnewline
\hline 

Star Name  & Alternative name  & Min dist  & Epoch  & Vrel  & r\_hel\tabularnewline
 &  & {[}pc{]}  & {[}Myr{]}  & {[}km s$^{-1}${]}  & {[}pc{]} \tabularnewline
\hline 
\endhead
\hline 
LP 646-17  & HIP 3757  & 0.04378  & -0.117935  & 207.96  & 3.175\tabularnewline
GJ 4274  & LP 820-12  & 0.41594  & -0.022642  & 316.28  & 0.611\tabularnewline
HD 761  & HIP 981  & 0.50883  & -6.674040  & 17.66  & 178.558\tabularnewline
GJ 3404 A  & TYC 151-860-1  & 0.55191  & -0.204269  & 61.21  & 5.498\tabularnewline
Wolf 28  & HIP 3829  & 0.65803  & -0.016006  & 266.49  & 0.432\tabularnewline
2MASS J07200325-0846499  &  & 0.90252  & -0.094773  & 60.63  & 2.552\tabularnewline
HD 135226  & BD-03 3748  & 0.91695  & -0.423887  & 73.23  & 11.408\tabularnewline
HD 232979  & HIP 21553  & 1.03139  & -0.281024  & 34.65  & 7.564\tabularnewline
TYC 5855-2215-1  &  & 1.04275  & -6.751784  & 39.20  & 180.616\tabularnewline
2MASS J10433508+1213149  &  & 1.15503  & -0.053937  & 266.61  & 1.453\tabularnewline
$\alpha$ Centauri B  & HIP 71681  & 1.25515  & 0.000000  & 35.39  & 0.001\tabularnewline
Proxima Centauri  & HIP 70890  & 1.30223  & 0.000000  & 37.09  & 0.001\tabularnewline
$\alpha$ Centauri A  & HIP 71683  & 1.32516  & 0.000000  & 35.03  & 0.001\tabularnewline
$\alpha$ Centauri AB  &  & 1.34803  & 0.000000  & 36.57  & 0.001\tabularnewline
UCAC4 535-065571  &  & 1.42794  & -2.140431  & 5.36  & 57.564\tabularnewline
GJ 358  & HIP 47425  & 1.54383  & -0.073742  & 122.26  & 1.986\tabularnewline
GJ 793  & HIP 101180  & 1.67110  & -0.236802  & 32.72  & 6.374\tabularnewline
Capella  & HIP 24608  & 1.75796  & -0.485275  & 25.83  & 13.060\tabularnewline
GJ 9603  & HIP 86916  & 1.79609  & -0.451911  & 43.43  & 12.163\tabularnewline
Barnard's star  & HIP 87937  & 1.82274  & 0.000000  & 134.83  & 0.001\tabularnewline
GJ 4063  & TYC 3109-1699-1  & 1.82991  & -0.175405  & 40.07  & 4.722\tabularnewline
GJ 195 A  & Capella H  & 1.84076  & -0.410834  & 32.61  & 11.057\tabularnewline
HD 200325  & HIP 103749  & 1.85997  & -4.379190  & 12.06  & 117.531\tabularnewline
G 208-45  & GJ 1245 B  & 1.94326  & -0.119017  & 33.73  & 3.204\tabularnewline
HZ 10  & WD 0407+179  & 1.97167  & -0.545641  & 68.15  & 14.685\tabularnewline
HD 8671  & HIP 6711  & 1.97438  & -1.106171  & 37.96  & 29.765\tabularnewline
BD+31 637  & TYC 2355-291-1  & 2.02908  & -5.090162  & 26.82  & 136.492\tabularnewline
GJ 9492  & HIP 71898  & 2.03615  & -0.282622  & 37.33  & 7.607\tabularnewline
L 923-22  & GJ 754.1 B  & 2.05851  & -0.176115  & 59.26  & 4.741\tabularnewline
HD 24546  & HIP 18453  & 2.06574  & -0.907453  & 40.07  & 24.419\tabularnewline
G 208-44  & GJ 1245 A  & 2.07612  & -0.122176  & 31.54  & 3.289\tabularnewline
LP 160-22  &  & 2.09918  & -0.526624  & 34.54  & 14.173\tabularnewline
HIP 34603  & GJ 268  & 2.10932  & -0.188072  & 31.99  & 5.063\tabularnewline
2MASS J18212815+1414010  &  & 2.11724  & -0.251783  & 35.39  & 6.777\tabularnewline
GJ 65 B  &  & 2.15456  & -0.039837  & 34.10  & 1.073\tabularnewline
GJ 909 A  & HIP 117712  & 2.17334  & -0.521136  & 20.37  & 14.025\tabularnewline
Wolf 359  & GJ 406  & 2.20907  & -0.027371  & 30.60  & 0.738\tabularnewline
GJ 876  & HIP 113020  & 2.24392  & -0.817099  & 5.07  & 21.989\tabularnewline
GJ 3376 A  & HIP 28267  & 2.24657  & -0.198471  & 121.88  & 5.342\tabularnewline
Kapteyn's star  & HIP 24186  & 2.28553  & -0.010651  & 270.91  & 0.288\tabularnewline
TYC 8470-213-1  &  & 2.30049  & -0.172274  & 229.85  & 4.637\tabularnewline
APMPM J0237-5928  &  & 2.30106  & -0.097524  & 97.49  & 2.626\tabularnewline
GJ 3287  &  & 2.31517  & -1.072967  & 20.63  & 28.872\tabularnewline
G 203-47  & HIP 83945  & 2.32770  & -0.093300  & 78.09  & 2.512\tabularnewline
HD 91962  & HIP 51966  & 2.36104  & -3.529085  & 11.12  & 94.802\tabularnewline
GJ 433.1  & HIP 56662  & 2.43579  & -0.126051  & 119.04  & 3.394\tabularnewline
HD 175726  & HIP 92984  & 2.46005  & -0.772211  & 33.25  & 20.781\tabularnewline
GJ 688  & HIP 86400  & 2.46025  & -0.255395  & 40.83  & 6.874\tabularnewline
HD 162826  & HIP 87382  & 2.51056  & -1.196101  & 27.79  & 32.183\tabularnewline
$\eta$ Casiopei B  & GJ 34 B  & 2.52756  & -0.205075  & 25.37  & 5.520\tabularnewline
HD 113376  & HIP 63797  & 2.52972  & -2.856293  & 40.47  & 76.776\tabularnewline
GJ 1095  & HIP 35136  & 2.54526  & -0.197668  & 81.92  & 5.321\tabularnewline
GJ 411  & HIP 54035  & 2.54679  & 0.000000  & 93.39  & 0.001\tabularnewline
HD 201671  & HIP 104539  & 2.55447  & -10.508025  & 10.14  & 279.312\tabularnewline
2MASSI J1835379+325954  &  & 2.58023  & -0.122667  & 39.52  & 3.302\tabularnewline
2MASS J05565722+1144333  &  & 2.60458  & -0.085689  & 144.85  & 2.307\tabularnewline
Aldebaran  & HIP 21421  & 2.60688  & -0.492455  & 40.82  & 13.254\tabularnewline
Sirius  & HIP 32349  & 2.63821  & 0.000000  & 31.91  & 0.001\tabularnewline
GJ 9105 C  & HD 18143C  & 2.67215  & -0.534272  & 28.67  & 14.379\tabularnewline
GJ 725 B  & HIP 91772  & 2.70964  & -0.058137  & 39.30  & 1.566\tabularnewline
CD-56 1032  & HIP 22738  & 2.75803  & -0.613484  & 17.28  & 16.510\tabularnewline
GJ 725 A  & HIP 91768  & 2.77896  & -0.058064  & 36.82  & 1.564\tabularnewline
GJ 54.1  & HIP 5643  & 2.79505  & -0.070215  & 32.52  & 1.891\tabularnewline
{[}D75b{]} Star 1  & G 217-32  & 2.82886  & -0.790870  & 17.31  & 21.283\tabularnewline
GJ 752  & HIP 94761  & 2.83223  & -0.070143  & 66.46  & 1.889\tabularnewline
GJ 15 A  & HIP 1475  & 2.87070  & -0.048115  & 39.20  & 1.296\tabularnewline
LP 412-31  &  & 2.87170  & -0.357225  & 38.54  & 9.615\tabularnewline
GJ 644  & HIP 82817  & 2.87451  & -0.146364  & 36.53  & 3.940\tabularnewline
TYC 7693-1161-1  &  & 2.90315  & -0.734088  & 47.41  & 19.755\tabularnewline
$\eta$ Casiopei  & HIP 3821  & 2.90499  & -0.212720  & 23.37  & 5.726\tabularnewline
GJ 729  & HIP 92403  & 2.91228  & -0.027389  & 21.15  & 0.738\tabularnewline
GJ 15 B  & HD 1326B  & 2.91483  & -0.048154  & 38.84  & 1.297\tabularnewline
TYC 5325-1808-1  &  & 2.93291  & -10.121533  & 7.94  & 269.220\tabularnewline
HD 317657  & TYC 7375-47-1  & 2.93350  & -0.408810  & 327.01  & 11.003\tabularnewline
Teegarden's star  & GAT 1370  & 2.94180  & -0.027297  & 86.90  & 0.736\tabularnewline
GJ 376 B  & HD 86728B  & 2.95912  & -0.255745  & 53.14  & 6.884\tabularnewline
TYC 5172-2349-1  &  & 3.00558  & -4.680653  & 35.14  & 125.576\tabularnewline
GJ 702  & HIP 88601  & 3.01970  & -0.218690  & 19.04  & 5.887\tabularnewline
Ruiz 207-61  &  & 3.09432  & -0.146225  & 89.76  & 3.936\tabularnewline
GJ 159.1  & HG 8-7  & 3.09673  & -0.329203  & 79.37  & 8.861\tabularnewline
HD 18768  & HIP 14181  & 3.09713  & -0.474439  & 99.02  & 12.769\tabularnewline
GJ 213  & HIP 26857  & 3.10918  & -0.048199  & 104.39  & 1.298\tabularnewline
GJ 644 C  &  & 3.12047  & -0.147195  & 37.54  & 3.963\tabularnewline
GJ 752 B  & VB 10  & 3.12731  & -0.085166  & 61.00  & 2.293\tabularnewline
GJ 905  & Ross 248  & 3.15691  & 0.000000  & 68.58  & 0.001\tabularnewline
GJ 160.1  & HIP 19255  & 3.17841  & -0.872863  & 22.35  & 23.489\tabularnewline
$\delta$ Capricorni  & HIP 107556  & 3.20613  & -1.557216  & 7.20  & 41.893\tabularnewline
HD 172051  & HIP 91438  & 3.21372  & -0.230299  & 52.90  & 6.199\tabularnewline
$\epsilon$ Eridani  & HIP 16537  & 3.21684  & 0.000000  & 33.32  & 0.001\tabularnewline
GJ 701  & HIP 88574  & 3.24097  & -0.121593  & 58.79  & 3.274\tabularnewline
GJ 643  & HIP 82809  & 3.25322  & -0.147410  & 37.72  & 3.968\tabularnewline
GJ 887  & HIP 114046  & 3.26819  & -0.002325  & 93.26  & 0.064\tabularnewline
LHS 1817  & LP 86-173  & 3.27213  & -0.097496  & 134.53  & 2.625\tabularnewline
Wolf 1059  & LHS 5131  & 3.27918  & -0.090442  & 258.71  & 2.435\tabularnewline
GJ 109  & HIP 12781  & 3.33010  & -0.212547  & 29.78  & 5.721\tabularnewline
GJ 447  & HIP 57548  & 3.38087  & 0.000000  & 47.70  & 0.001\tabularnewline
2MASS J16452211-1319516  &  & 3.38729  & -0.174101  & 45.74  & 4.687\tabularnewline
HD 132730  & TYC 8681-841-1  & 3.39323  & -3.989147  & 15.80  & 107.110\tabularnewline
GJ 866  & LP 820-64  & 3.40568  & 0.000000  & 78.04  & 0.001\tabularnewline
GJ 745 A  & HIP 93873  & 3.41696  & -0.121135  & 62.74  & 3.261\tabularnewline
G 123-45  &  & 3.42044  & -0.463265  & 22.31  & 12.468\tabularnewline
HD 174153  & HIP 92519  & 3.43893  & -0.646559  & 77.15  & 17.400\tabularnewline
GJ 3988  & LHS 3262  & 3.44475  & -0.119955  & 70.73  & 3.229\tabularnewline
GJ 663 A  & TYC 6820-326-1  & 3.44561  & -0.261330  & 14.76  & 7.034\tabularnewline
GJ 745 B  & HIP 93899  & 3.45640  & -0.121402  & 62.52  & 3.268\tabularnewline
GJ 170  & LHS 1674  & 3.45761  & -0.403927  & 25.86  & 10.871\tabularnewline
BD+50 860B  & TYC 3339-1311-1  & 3.47850  & -1.406654  & 26.18  & 37.845\tabularnewline
61 Cygni A  & HIP 104214  & 3.48546  & 0.000000  & 88.99  & 0.001\tabularnewline
61 Cygni B  & HIP 104217  & 3.49692  & 0.000000  & 86.54  & 0.001\tabularnewline
\end{longtable}

\newpage{}
{\footnotesize 
{\setlength{\tabcolsep}{2.5pt} 
\begin{longtable}{llrrrrrrl}
\caption{\label{A2} Final list of 57 stars or star systems used as perturbers
in our final dynamical model of 'Oumuamua's motion.  Heliocentric
Galactic velocities are in parsecs per Myr. To obtain velocities in
km$^{-}1$, each component must be divided by 1.0227. Stars are presented
here in the same sequence as in Table A3. \textnormal{In the last column we include references for all data used to produce this table. For each star or stellar system we present sources of: positions, proper motions, parallaxes, radial velocities and masses in this order. In some cases of multiple systems we present individual member references connected with a plus sign.}}
\tabularnewline
\hline 
Star name  & Mass  & \multicolumn{1}{c}{X}  & \multicolumn{1}{c}{Y}  & \multicolumn{1}{c}{Z}  & \multicolumn{1}{c}{Vx}  & \multicolumn{1}{c}{Vy}  &\multicolumn{1}{c}{Vz} & Ref\tabularnewline
 & M$_{\odot}$  &\multicolumn{1}{c}{ {[}pc{]}}  & \multicolumn{1}{c}{ {[}pc{]}}   & \multicolumn{1}{c}{ {[}pc{]}}   & {[}pc Myr$^{-1}${]}  & {[}pc Myr$^{-1}${]}  & {[}pc Myr$^{-1}${]} & \tabularnewline
\endfirsthead
\hline 
\caption{continued.}
\tabularnewline
\hline 
Star names  & Mass  & X  & Y  & Z  & Vx  & Vy  & Vz & Ref\tabularnewline
 & M$_{\odot}$  & {[}pc{]}  & {[}pc{]}  & {[}pc{]}  & {[}pc Myr$^{-1}${]}  & {[}pc Myr$^{-1}${]}  & {[}pc Myr$^{-1}${]} & \tabularnewline
\endhead
\hline 
HIP 3757  & 0.4  & -4.869905  & 8.163363  & -23.203674  & -52.989230  & 46.668895  & -204.591445  & 1,1,1,2,3     \tabularnewline
GJ 4274  & 0.14  & 3.223762  & 2.957145  & -6.039148  & 141.140365  & 94.217546  & -267.797888  & 4,5,6,2,7     \tabularnewline
TYC 151-860-1  & 0.23  & -11.260623  & -6.162569  & -0.094524  & -66.113087  & -53.790236  & -5.867103  &  5,8,9,10,11    \tabularnewline
HIP 3829, Wolf 28  & 0.68  & -1.210292  & 1.945959  & -3.594159  & -72.175698  & 69.395114  & -257.111463  & 12,12,12,13,14     \tabularnewline
2MASS J07200325-0846499  & 0.08  & -4.426203  & -4.074046  & 0.317199  & -60.908353  & -59.344984  & 1.872151  & 4,15,15,16,17   \tabularnewline
TYC 5009-283-1  & 1.0  & 23.021112  & -1.443553  & 21.958699  & 43.238409  & -24.467213  & 42.944681  & 18,18,18,19,2     \tabularnewline
HIP 21553  & 0.6  & -8.799950  & 4.438152  & 0.669586  & -43.309867  & -8.021714  & -1.980970  & 1,1,1,20,21     \tabularnewline
2MASS J10433508+1213149  & 0.08  & -4.833887  & -6.430379  & 12.182001  & -83.878243  & -136.754016  & 229.107657  & 4,22,22,22,23\tabularnewline
$\alpha$Cent AB+Proxima system  & 2.17  & 0.935517  & -0.893740  & 0.017400  & -29.182899  & 1.017810  & 12.535209  & 12,12,12+24,25+26+27,28     \tabularnewline
UCAC4 535-065571  & 0.205  & 8.142407  & 6.867308  & 4.863749  & -8.368095  & -19.233016  & -5.616024  & 4,5,9,10,21     \tabularnewline
HIP 47425  & 0.34  & -0.270092  & -9.499987  & 1.460452  & -33.489157  & -143.256893  & 18.063286  & 1,1,1,29,30     \tabularnewline
HIP 101180  & 0.39  & -1.342284  & 7.662756  & 2.093016  & -20.645740  & 9.234058  & -5.321937  & 1,1,1,31,32     \tabularnewline
HIP 24608  & 5.06  & -12.482300  & 3.914475  & 1.044814  & -36.754471  & -14.635561  & -9.310078  & 12,12,12,33,34     \tabularnewline
HIP 86916  & 0.5  & 4.967614  & 16.744254  & 10.225582  & 2.385149  & 14.079262  & 12.221230  & 1,1,1,35,32     \tabularnewline
TYC 3109-1699-1  & 0.2  & 2.479393  & 6.336940  & 2.491140  & 10.607473  & 10.433390  & 0.444911  & 5,8,36,37,38     \tabularnewline
HIP 87937  & 0.16  & 1.516300  & 0.911399  & 0.443149  & -144.200647  & 4.701043  & 18.610843  & 12,12,12,31,41     \tabularnewline
HZ 10  & 1.0  & -34.764964  & 2.772228  & -15.508177  & -75.377293  & -14.473130  & -34.971842  & 5,5,39,40,42     \tabularnewline
HIP 6711  & 1.39  & -25.879842  & 31.269062  & -13.945369  & -35.841132  & 5.440825  & -18.856965  & 1,1,1,13,43     \tabularnewline
GJ 1245 ABC system  & 0.28  & 0.868629  & 4.410703  & 0.672563  & 6.129215  & 4.675136  & -11.727727  &  5,5+44,45,10+46,47\tabularnewline
HIP 71898  & 0.48  & -2.134723  & 7.105483  & 8.076950  & -12.242474  & 1.061814  & 21.798225  & 1,1,1,31,50     \tabularnewline
L 923-22  & 0.1  & 9.336853  & 5.513232  & -1.784881  & 38.014847  & 11.743817  & -7.341221  & 48,5,49,37,30     \tabularnewline
HIP 34603  & 0.33  & -5.917432  & 0.108014  & 2.143010  & -44.405820  & -22.043392  & -7.659702  & 12,12,12,33,51     \tabularnewline
GJ 65 AB  & 0.21  & -0.697183  & 0.119872  & -2.527542  & -44.334435  & -18.602954  & -19.451010  & 52,52,52,53+54,55     \tabularnewline
HIP 117712  & 1.09  & -5.173173  & 9.260300  & 2.469058  & -18.194848  & -4.381304  & -0.938894  & 12,12,12,13,56     \tabularnewline
GJ 406  & 0.09  & -0.583077  & -1.198362  & 1.984705  & -28.448478  & -48.653982  & -13.935499  & 4,5,45,37,45     \tabularnewline
HIP 28267  & 1.3  & -20.646959  & -11.843636  & -5.640298  & -110.785271  & -92.744647  & -36.941301  &  1,1,1,31,57    \tabularnewline
HIP 113020  & 0.334  & 1.441102  & 1.905640  & -4.034279  & -12.583839  & -20.164310  & -12.213344  & 12,12,12,37,58     \tabularnewline
TYC 8470-213-1  & 1.0  & 15.179662  & -15.745125  & -34.620438  & 82.441710  & -102.422426  & -208.063306  & 1,1,1,2,2     \tabularnewline
HIP 24186  & 0.39  & -1.057028  & -2.982685  & -2.299061  & 19.991822  & -294.248666  & -54.060419  & 12,12,12,31,59     \tabularnewline
APMPM J0237-5928  & 0.22  & 0.989962  & -5.626783  & -7.766258  & -23.371954  & -73.481326  & -93.028655  & 5,60,61,2,62     \tabularnewline
HIP 83945  & 0.266  & 2.166005  & 5.597234  & 4.406165  & 26.034702  & 41.869697  & 19.607791  & 12,12,12,37,47     \tabularnewline
HIP 56662  & 0.935  & -4.247950  & -1.484937  & 15.153736  & -41.484216  & -15.924967  & 109.922558  & 12,12,12,63,64     \tabularnewline
HIP 86400  & 0.85  & 9.274425  & 4.883523  & 3.336740  & 17.935902  & -0.693717  & 11.375820  & 12,12,12,19,65     \tabularnewline
HIP 35136  & 0.9  & -15.269627  & 2.598365  & 6.739659  & -81.650214  & -1.684617  & 33.045457  & 12,12,12,31,66     \tabularnewline
HIP 54035  & 0.46  & -1.054711  & -0.094472  & 2.316283  & 47.197616  & -54.908745  & -75.981741  &  12,12,12,31,67    \tabularnewline
2MASSI J1835379+325954  & 0.07  & 2.435869  & 4.513765  & 1.611691  & 21.323286  & -0.059501  & -3.403988  & 4,68,9,69,9     \tabularnewline
2MASS J05565722+1144333  & 0.15  & -12.556104  & -3.645129  & -1.479015  & -148.572944  & -78.990706  & 0.269334  &  4,44,9,70,71\tabularnewline
HIP 32349  & 2.99  & -1.815217  & -1.874932  & -0.379090  & 13.087100  & -2.296652  & -12.179029  & 12,12,12,13,72     \tabularnewline
GJ 725 AB system  & 0.58  & 0.039863  & 3.202252  & 1.441291  & -25.325856  & -11.833516  & 26.619384  &  1,1,1,31,47    \tabularnewline
HIP 22738  & 0.7  & -1.192678  & -8.579194  & -6.955349  & -9.646365  & -35.215095  & -20.407536  & 12,12,12,29,50     \tabularnewline
HIP 5643  & 0.13  & -0.692360  & 0.466957  & -3.594156  & -29.284995  & 0.424845  & -23.796103  &  12,12,12,31,30    \tabularnewline
GJ 752 AB system  & 0.55  & 4.481805  & 3.820433  & -0.338091  & 53.884411  & -8.729523  & -5.101046  &  4+1,73+1,73+1,10+20,47 \tabularnewline
GJ 15 AB system  & 0.65  & -1.519073  & 3.023091  & -1.128502  & -50.153963  & -12.747165  & -3.667871  & 1+4,1+44,1+74,31+37,21 \tabularnewline
HIP 92403  & 0.17  & 2.855273  & 0.650863  & -0.493625  & -12.182465  & -1.279999  & -7.549572  & 12,12,12,31,21     \tabularnewline
Teegarden's star  & 0.08  & -2.883244  & 1.034454  & -2.310472  & -66.768384  & -73.067253  & -56.387473  & 4,44,61,10,21     \tabularnewline
HIP 88601  & 1.62  & 4.320600  & 2.483796  & 1.001895  & 6.169135  & -19.189299  & -14.679349  &  12,12,12,75,76    \tabularnewline
Ruiz 207-61  & 0.08  & -2.466202  & -13.392706  & -4.316815  & -44.855241  & -106.443188  & -26.577086  & 4,68,45,77,78     \tabularnewline
HIP 26857  & 0.18  & -5.590121  & -1.363571  & -0.933193  & -91.980164  & -91.361241  & 8.418275  & 12,12,12,79,47     \tabularnewline
Ross 248  & 0.12  & -1.032642  & 2.838561  & -0.920011  & 33.153204  & -76.262790  & 0.284634  &  4,80,80,37,21    \tabularnewline
HIP 91438  & 0.9  & 12.669348  & 2.882727  & -1.537885  & 38.206785  & -2.328500  & -4.403618  &  12,12,12,19,81    \tabularnewline
$\delta$Capricorni  & 2.06  & 6.511185  & 5.075845  & -8.523903  & -7.436483  & -18.343120  & -11.762829  & 12,12,12,33,82     \tabularnewline
HIP 16537  & 0.85  & -2.098618  & -0.549497  & -2.374208  & -3.862226  & 7.417522  & -21.063553  & 12,12,12,31,83     \tabularnewline
HIP 88574  & 0.5  & 6.950725  & 3.231218  & 1.200757  & 33.830913  & 14.644326  & -19.358860  & 12,12,12,31,30     \tabularnewline
HIP 114046  & 0.53  & 1.299173  & 0.211052  & -2.999850  & -96.274433  & -13.614020  & -51.028239  &  12,12,12,84,30    \tabularnewline
HIP 57548  & 0.16  & 0.004397  & -1.712804  & 2.914646  & 18.364701  & 5.194591  & -33.949087  & 1,1,1,37,21     \tabularnewline
GJ 866 ABC system  & 0.30  & 1.249826  & 1.390358  & -2.847040  & -69.843346  & -1.090148  & 42.215336  & 4,85,86,25,87     \tabularnewline
61 Cyg AB system  & 1.33  & 0.464345  & 3.442752  & -0.354214  & -95.544501  & -55.395340  & -8.848684 & 12,12,12,31,88 \tabularnewline
\end{longtable}
}}

\tablebib{
( 1)~\citet{ 2016A&A...595A...2G};
( 2)~\citet{ 2013AJ....146..134K};
( 3)~\citet{ 2014MNRAS.443.2561G};
( 4)~\citet{ 2003yCat.2246....0C};
( 5)~\citet{ 2012yCat.1322....0Z};
( 6)~\citet{ 2017ARep...61..883B};
( 7)~\citet{ 2013A&A...549A.109B};
( 8)~\citet{ 2000A&A...355L..27H};
( 9)~\citet{ 2014ApJ...784..156D};
(10)~\citet{ 2014AJ....147...20N};
(11)~\citet{ 2013A&A...551A..36N};
(12)~\citet{ vanleeuwen:2007};
(13)~\citet{ Gontcharov:2006};
(14)~\citet{ 2012ApJS..199...29G};
(15)~\citet{ 2012ApJ...752...56F};
(16)~\citet{ 2009yCat.2294....0A};
(17)~\citet{ 2017yCat.4034....0H};
(18)~\citet{ Gaia_mission:2016};
(19)~\citet{ RAVE-5:2017};
(20)~\citet{ 2013A&A...552A..64S};
(21)~\citet{ 2017ApJ...834...85N};
(22)~\citet{ 2015AJ....149..104B};
(23)~\citet{ 2015AJ....150..180B};
(24)~\citet{ 2012ApJS..201...19D};
(25)~\citet{ 1953GCRV..C......0W};
(26)~\citet{ 2005ApJS..159..141V};
(27)~\citet{ 2006A&A...460..695T};
(28)~\citet{ 2017A&A...598L...7K};
(29)~\citet{ 2001A&A...379..634G};
(30)~\citet{ 2017A&A...600A..13A};
(31)~\citet{ 2002ApJS..141..503N};
(32)~\citet{ 2017A&A...598A..26P};
(33)~\citet{ 2004MNRAS.349.1069K};
(34)~\citet{ 2015ApJ...807...26T};
(35)~\citet{ 2007AN....328..889K};
(36)~\citet{ 2002AJ....123.2806R};
(37)~\citet{ 2015ApJ...802L..10T};
(38)~\citet{ 2014ApJ...791...54G};
(39)~\citet{ 1995GCTP..C......0V};
(40)~\citet{ 1967ApJ...149..283G};
(41)~\citet{ 2012ApJ...747..144M};
(42)~\citet{ 2015MNRAS.454.2787G};
(43)~\citet{ 2012ApJ...756...46R};
(44)~\citet{ 2005AJ....129.1483L};
(45)~\citet{ 2009ApJ...704..975J};
(46)~\citet{ 2010MNRAS.407.2269M};
(47)~\citet{ 2015ApJ...804...64M};
(48)~\citet{ 2003yCat.1289....0Z};
(49)~\citet{ 1980AJ.....85..454H};
(50)~\citet{ 2015MNRAS.449.2618W};
(51)~\citet{ 2012ApJ...760...55B};
(52)~\citet{ 2015AJ....149....5W};
(53)~\citet{ 2001MNRAS.328...45M};
(54)~\citet{ 2016AJ....152..141B};
(55)~\citet{ 2016A&A...593A.127K};
(56)~\citet{ 2015A&A...574A...6A};
(57)~\citet{ 2007A&A...474..273E};
(58)~\citet{ 2010A&A...511A..21C};
(59)~\citet{ 2014MNRAS.443L..89A};
(60)~\citet{ 2000A&A...353..958S};
(61)~\citet{ 2006AJ....132.2360H};
(62)~\citet{ 1999A&A...345L..55S};
(63)~\citet{ 2013AJ....146...34Z};
(64)~\citet{ 2017MNRAS.465.2849T};
(65)~\citet{ 2013AJ....145...41K};
(66)~\citet{ 2011A&A...530A.138C};
(67)~\citet{ 2003A&A...397L...5S};
(68)~\citet{ 2007AJ....133.2258S};
(69)~\citet{ 2012AJ....144...99D};
(70)~\citet{ 2003AJ....125.1598L};
(71)~\citet{ Newton-et-al:2016};
(72)~\citet{ 2017ApJ...840...70B};
(73)~\citet{ 1992AJ....103..638M};
(74)~\citet{ 2014ApJ...784..156D};
(75)~\citet{ 2004A&A...424..727P};
(76)~\citet{ 1998A&A...338..455F};
(77)~\citet{ 2015ApJS..220...18B};
(78)~\citet{ 2015ApJ...810..158F};
(79)~\citet{ 2011AAS...21743412C};
(80)~\citet{ 2008AJ....136..452G};
(81)~\citet{ 2017ApJ...836..139F};
(82)~\citet{ 2014A&A...572A..91T};
(83)~\citet{ 2014MNRAS.445.2223P};
(84)~\citet{ 2015A&A...573A.126D};
(85)~\citet{ 2003ApJ...582.1011S};
(86)~\citet{ 2010A&ARv..18...67T};
(87)~\citet{ 2000A&A...364..665S};
(88)~\citet{ 2008A&A...488..667K};
}


\vspace{2cm}

\begin{longtable}{lcccc}
\caption{\label{A3}Minimal distances between the 10\,000 clones of 'Oumuamua
and 57 stars included in our research. Here we present variation intervals
roughly equivalent to the $\pm$4$\sigma$ deviations, obtained from
among the 10\.{0}00 encounters. In the last column we present a heliocentric
distance for this event. Epochs are in the past, but minuses are omitted
here for sake of clarity. Proximity epochs equal to zero mean that
the star's closest position is at the beginning of the backward numerical
integration (but they are still acting as perturbers). }
\tabularnewline
\hline 
\hline 
Star name  & min dist  & epoch  & rel velocity  & dist\_hel\tabularnewline
 & {[}pc{]}  & {[}Myr{]}  & {[}km s$^{-1}${]}  & {[}pc{]} \tabularnewline
\endfirsthead
\caption{continued.}
\tabularnewline
\hline 
\hline 
Star name  & min dist  & epoch  & rel velocity  & dist\_hel\tabularnewline
 & {[}pc{]}  & {[}Myr{]}  & {[}km s$^{-1}${]}  & {[}pc{]} \tabularnewline
\hline 
\endhead
\hline 
HIP 3757  & 0.042 : 0.044 : 0.047  & 0.1179 : 0.1179 : 0.1180  & 208.00 : 207.96 : 207.92  & 3.17 : 3.17 : 3.18 \tabularnewline
GJ 4274  & 0.411 : 0.412 : 0.413  & 0.0227 : 0.0227 : 0.0227  & 316.28 : 316.27 : 316.27  & 0.61 : 0.61 : 0.61 \tabularnewline
TYC 151-860-1  & 0.543 : 0.551 : 0.559  & 0.2044 : 0.2046 : 0.2047  & 61.24 : 61.21 : 61.17  & 5.50 : 5.51 : 5.52 \tabularnewline
HIP 3829, Wolf 28  & 0.644 : 0.644 : 0.645  & 0.0152 : 0.0152 : 0.0152  & 266.50 : 266.49 : 266.52  & 0.41 : 0.41 : 0.41 \tabularnewline
2MASS J07200325-0846499  & 0.899 : 0.900 : 0.901  & 0.0953 : 0.0953 : 0.0954  & 60.65 : 60.63 : 60.60  & 2.56 : 2.57 : 2.57 \tabularnewline
TYC 5009-283-1  & 0.920 : 0.924 : 0.928  & 0.4271 : 0.4270 : 0.4268  & 73.18 : 73.23 : 73.29  & 11.48 : 11.49 : 11.50 \tabularnewline
HIP 21553  & 1.013 : 1.023 : 1.033  & 0.2772 : 0.2774 : 0.2778  & 34.67 : 34.65 : 34.62  & 7.45 : 7.47 : 7.49 \tabularnewline
2MASS J10433508+1213149  & 1.154 : 1.157 : 1.160  & 0.0540 : 0.0540 : 0.0540  & 266.60 : 266.62 : 266.63  & 1.45 : 1.45 : 1.46 \tabularnewline
$\alpha$Cen AB+Proxima system  & 1.294 : 1.294 : 1.294  & 0.0000 : 0.0000 : 0.0000  & 35.27 : 35.27 : 35.27  & 0.00 : 0.00 : 0.00 \tabularnewline
UCAC4 535-065571  & 1.372 : 1.427 : 1.484  & 2.1292 : 2.1395 : 2.1490  & 5.40 : 5.36 : 5.32  & 57.33 : 57.54 : 57.73 \tabularnewline
HIP 47425  & 1.517 : 1.519 : 1.521  & 0.0758 : 0.0758 : 0.0758  & 122.26 : 122.26 : 122.26  & 2.04 : 2.04 : 2.04 \tabularnewline
HIP 101180  & 1.659 : 1.671 : 1.683  & 0.2346 : 0.2345 : 0.2344  & 32.71 : 32.72 : 32.72  & 6.32 : 6.31 : 6.30 \tabularnewline
HIP 24608  & 1.732 : 1.748 : 1.764  & 0.4927 : 0.4927 : 0.4928  & 25.79 : 25.83 : 25.87  & 13.28 : 13.26 : 13.25 \tabularnewline
HIP 86916  & 1.790 : 1.794 : 1.799  & 0.4533 : 0.4533 : 0.4532  & 43.40 : 43.43 : 43.46  & 12.19 : 12.20 : 12.21 \tabularnewline
TYC 3109-1699-1  & 1.820 : 1.822 : 1.823  & 0.1705 : 0.1710 : 0.1715  & 40.04 : 40.07 : 40.10  & 4.58 : 4.60 : 4.62 \tabularnewline
HIP 87937  & 1.823 : 1.823 : 1.823  & 0.0000 : 0.0000 : 0.0000  & 134.80 : 134.83 : 134.86  & 0.00 : 0.00 : 0.00 \tabularnewline
HZ 10  & 1.964 : 1.975 : 1.986  & 0.5450 : 0.5449 : 0.5447  & 68.20 : 68.15 : 68.10  & 14.65 : 14.66 : 14.68 \tabularnewline
HIP 6711  & 1.958 : 1.976 : 1.995  & 1.1063 : 1.1068 : 1.1074  & 37.99 : 37.96 : 37.93  & 29.73 : 29.78 : 29.83 \tabularnewline
GJ 1245 ABC system  & 2.023 : 2.024 : 2.024  & 0.1238 : 0.1225 : 0.1229  & 32.31 : 32.33 : 32.35  & 3.33 : 3.30 : 3.31 \tabularnewline
HIP 71898  & 2.025 : 2.038 : 2.052  & 0.2844 : 0.2845 : 0.2848  & 37.30 : 37.33 : 37.36  & 7.64 : 7.66 : 7.67 \tabularnewline
L 923-22  & 2.051 : 2.056 : 2.060  & 0.1790 : 0.1769 : 0.1771  & 59.23 : 59.26 : 59.30  & 4.81 : 4.76 : 4.77 \tabularnewline
HIP 34603  & 2.091 : 2.098 : 2.105  & 0.1819 : 0.1818 : 0.1815  & 31.95 : 31.99 : 32.03  & 4.90 : 4.89 : 4.88 \tabularnewline
GJ 65 AB  & 2.151 : 2.152 : 2.153  & 0.0438 : 0.0438 : 0.0438  & 34.15 : 34.10 : 34.05  & 1.18 : 1.18 : 1.18 \tabularnewline
HIP 117712  & 2.138 : 2.166 : 2.194  & 0.5135 : 0.5135 : 0.5134  & 20.37 : 20.37 : 20.37  & 13.80 : 13.82 : 13.83 \tabularnewline
GJ 406  & 2.207 : 2.208 : 2.209  & 0.0295 : 0.0295 : 0.0295  & 30.57 : 30.60 : 30.63  & 0.80 : 0.80 : 0.79 \tabularnewline
HIP 28267  & 2.211 : 2.215 : 2.219  & 0.1951 : 0.1955 : 0.1958  & 121.92 : 121.88 : 121.84  & 5.25 : 5.26 : 5.28 \tabularnewline
HIP 113020  & 2.219 : 2.241 : 2.264  & 0.7902 : 0.7907 : 0.7917  & 5.04 : 5.07 : 5.11  & 21.29 : 21.28 : 21.28 \tabularnewline
TYC 8470-213-1  & 2.263 : 2.268 : 2.276  & 0.1738 : 0.1743 : 0.1746  & 229.86 : 229.85 : 229.83  & 4.67 : 4.69 : 4.71 \tabularnewline
HIP 24186  & 2.273 : 2.274 : 2.274  & 0.0115 : 0.0115 : 0.0115  & 270.91 : 270.91 : 270.90  & 0.31 : 0.31 : 0.31 \tabularnewline
APMPM J0237-5928  & 2.274 : 2.278 : 2.282  & 0.0954 : 0.0953 : 0.0953  & 97.45 : 97.49 : 97.52  & 2.57 : 2.57 : 2.56 \tabularnewline
HIP 83945  & 2.297 : 2.298 : 2.299  & 0.0889 : 0.0888 : 0.0888  & 78.12 : 78.09 : 78.05  & 2.40 : 2.39 : 2.39 \tabularnewline
HIP 56662  & 2.419 : 2.421 : 2.425  & 0.1282 : 0.1288 : 0.1293  & 119.01 : 119.04 : 119.06  & 3.45 : 3.47 : 3.49 \tabularnewline
HIP 86400  & 2.459 : 2.461 : 2.463  & 0.2552 : 0.2548 : 0.2544  & 40.88 : 40.83 : 40.79  & 6.88 : 6.86 : 6.84 \tabularnewline
HIP 35136  & 2.533 : 2.542 : 2.551  & 0.1990 : 0.1995 : 0.1998  & 81.94 : 81.92 : 81.91  & 5.35 : 5.37 : 5.38 \tabularnewline
HIP 54035  & 2.547 : 2.547 : 2.547  & 0.0000 : 0.0000 : 0.0000  & 93.39 : 93.39 : 93.39  & 0.00 : 0.00 : 0.00 \tabularnewline
2MASSI J1835379+325954  & 2.569 : 2.569 : 2.570  & 0.1173 : 0.1173 : 0.1162  & 39.48 : 39.52 : 39.55  & 3.15 : 3.16 : 3.13 \tabularnewline
2MASS J05565722+1144333  & 2.605 : 2.608 : 2.612  & 0.0854 : 0.0854 : 0.0854  & 144.89 : 144.85 : 144.81  & 2.30 : 2.30 : 2.30 \tabularnewline
HIP 32349  & 2.638 : 2.638 : 2.638  & 0.0000 : 0.0000 : 0.0000  & 31.89 : 31.91 : 31.94  & 0.00 : 0.00 : 0.00 \tabularnewline
GJ 725 AB system  & 2.747 : 2.748 : 2.748  & 0.0573 : 0.0573 : 0.0573  & 37.89 : 37.87 : 37.85  & 1.54 : 1.54 : 1.54 \tabularnewline
HIP 22738  & 2.740 : 2.758 : 2.776  & 0.6045 : 0.6042 : 0.6145  & 17.30 : 17.28 : 17.26  & 16.25 : 16.26 : 16.56 \tabularnewline
HIP 5643  & 2.792 : 2.794 : 2.796  & 0.0721 : 0.0721 : 0.0721  & 32.55 : 32.52 : 32.48  & 1.94 : 1.94 : 1.95 \tabularnewline
GJ 752 AB system  & 2.840 : 2.841 : 2.842  & 0.0775 : 0.0775 : 0.0775  & 65.63 : 65.67 : 65.71  & 2.08 : 2.09 : 2.09 \tabularnewline
GJ 15 AB system  & 2.875 : 2.876 : 2.876  & 0.0526 : 0.0526 : 0.0526  & 39.14 : 39.11 : 39.07  & 1.41 : 1.42 : 1.42 \tabularnewline
HIP 92403  & 2.911 : 2.912 : 2.913  & 0.0270 : 0.0270 : 0.0262  & 21.16 : 21.16 : 21.15  & 0.73 : 0.73 : 0.71 \tabularnewline
Teegarden's star  & 2.941 : 2.942 : 2.942  & 0.0277 : 0.0277 : 0.0277  & 86.95 : 86.90 : 86.85  & 0.75 : 0.75 : 0.75 \tabularnewline
HIP 88601  & 3.007 : 3.015 : 3.023  & 0.2105 : 0.2103 : 0.2102  & 19.06 : 19.04 : 19.01  & 5.67 : 5.66 : 5.65 \tabularnewline
Ruiz 207-61  & 3.047 : 3.052 : 3.052  & 0.1509 : 0.1538 : 0.1537  & 89.73 : 89.76 : 89.77  & 4.07 : 4.14 : 4.14 \tabularnewline
HIP 26857  & 3.101 : 3.103 : 3.105  & 0.0457 : 0.0458 : 0.0458  & 104.41 : 104.39 : 104.36  & 1.23 : 1.23 : 1.24 \tabularnewline
Ross 248  & 3.157 : 3.157 : 3.157  & 0.0000 : 0.0000 : 0.0000  & 68.56 : 68.58 : 68.61  & 0.00 : 0.00 : 0.00 \tabularnewline
HIP 91438  & 3.202 : 3.206 : 3.210  & 0.2344 : 0.2345 : 0.2346  & 52.86 : 52.90 : 52.94  & 6.30 : 6.31 : 6.32 \tabularnewline
$\delta$Capricorni  & 3.168 : 3.212 : 3.259  & 1.5214 : 1.5259 : 1.5317  & 7.20 : 7.20 : 7.21  & 40.88 : 41.05 : 41.26 \tabularnewline
HIP 16537  & 3.217 : 3.217 : 3.217  & 0.0000 : 0.0000 : 0.0000  & 33.32 : 33.32 : 33.32  & 0.00 : 0.00 : 0.00 \tabularnewline
HIP 88574  & 3.225 : 3.231 : 3.236  & 0.1173 : 0.1173 : 0.1173  & 58.81 : 58.79 : 58.76  & 3.16 : 3.16 : 3.15 \tabularnewline
HIP 114046  & 3.268 : 3.268 : 3.268  & 0.0023 : 0.0023 : 0.0023  & 93.31 : 93.26 : 93.21  & 0.06 : 0.06 : 0.06 \tabularnewline
HIP 57548  & 3.381 : 3.381 : 3.381  & 0.0000 : 0.0000 : 0.0000  & 47.71 : 47.70 : 47.70  & 0.00 : 0.00 : 0.00 \tabularnewline
GJ 866 ABC system  & 3.406 : 3.406 : 3.406  & 0.0000 : 0.0000 : 0.0000  & 78.05 : 78.04 : 78.04  & 0.00 : 0.00 : 0.00 \tabularnewline
61 Cyg AB system  & 3.491 : 3.491 : 3.491  & 0.0000 : 0.0000 : 0.0000  & 87.79 : 87.83 : 87.86  & 0.00 : 0.00 : 0.00 \tabularnewline
\end{longtable}


\begin{table}
\caption{\label{geometries} Examples of the geometry of 'Oumuamua encounters with selected stars.  We use here a heliocentric, non-rotating, right-handed rectangular frame. The XY plane is parallel to the Galactic disk plane and the OX axis is directed to the Galactic Centre at the beginning of the calculation.   Green lines depict the 'Oumuamua motion while the red ones show the star trajectory. Open circles mark the starting positions of 'Oumuamua and the star.}

\vspace{-0.7cm}

\begin{tabular}{ p{8.5cm} p{8.5cm}}
\begin{center}	
\includegraphics[angle=270, width=0.42\textwidth]{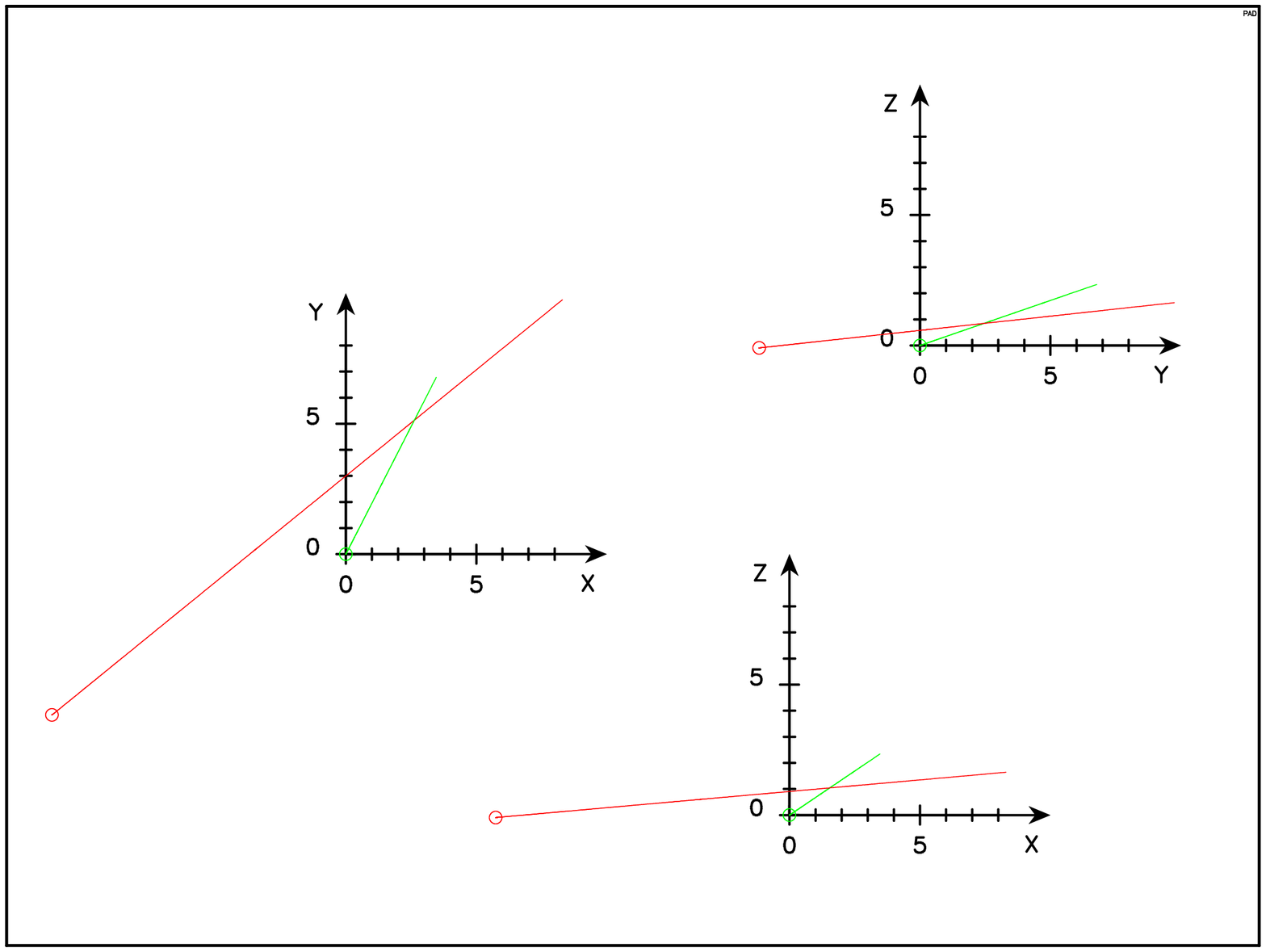}
\label{fig:startrek22g108-21a}
\end{center}	
&  
\begin{center}	
\includegraphics[angle=270, width=0.42\textwidth]{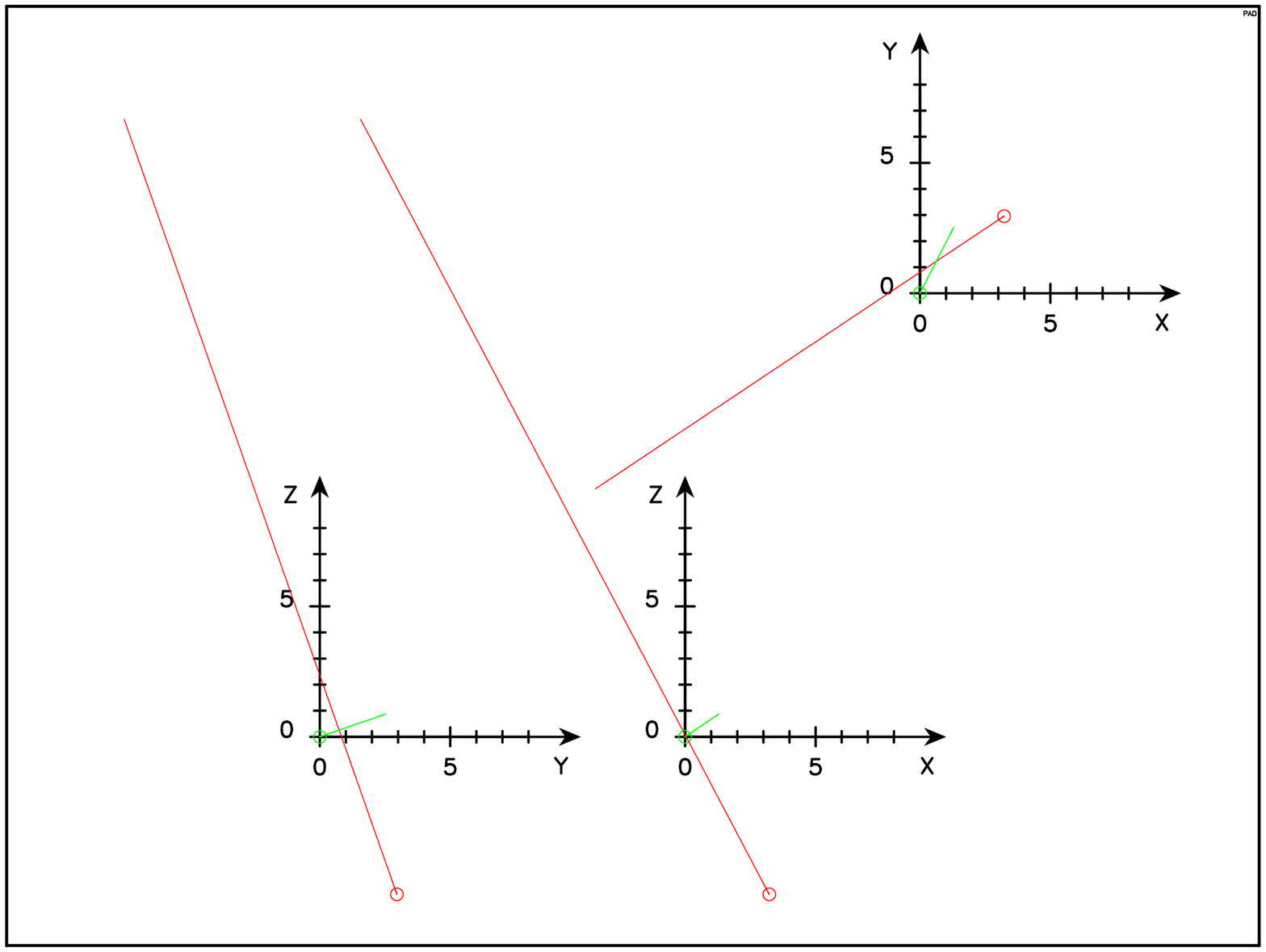}
\label{fig:startrek22gj4274a}
\end{center}
\\ 

Fig.~A.1. The geometry of the encounter of 'Oumuamua with the star G\,108-21 0.2 Myr ago. Depicted is 0.3 Myr of their motion.

&
 
Fig.~A.2. The geometry of the encounter of 'Oumuamua with the star GJ\,4274 23 kyr ago. 112 kyr of motion of these bodies is shown here.

\\
	\begin{center}
\includegraphics[angle=270, width=0.42\textwidth]{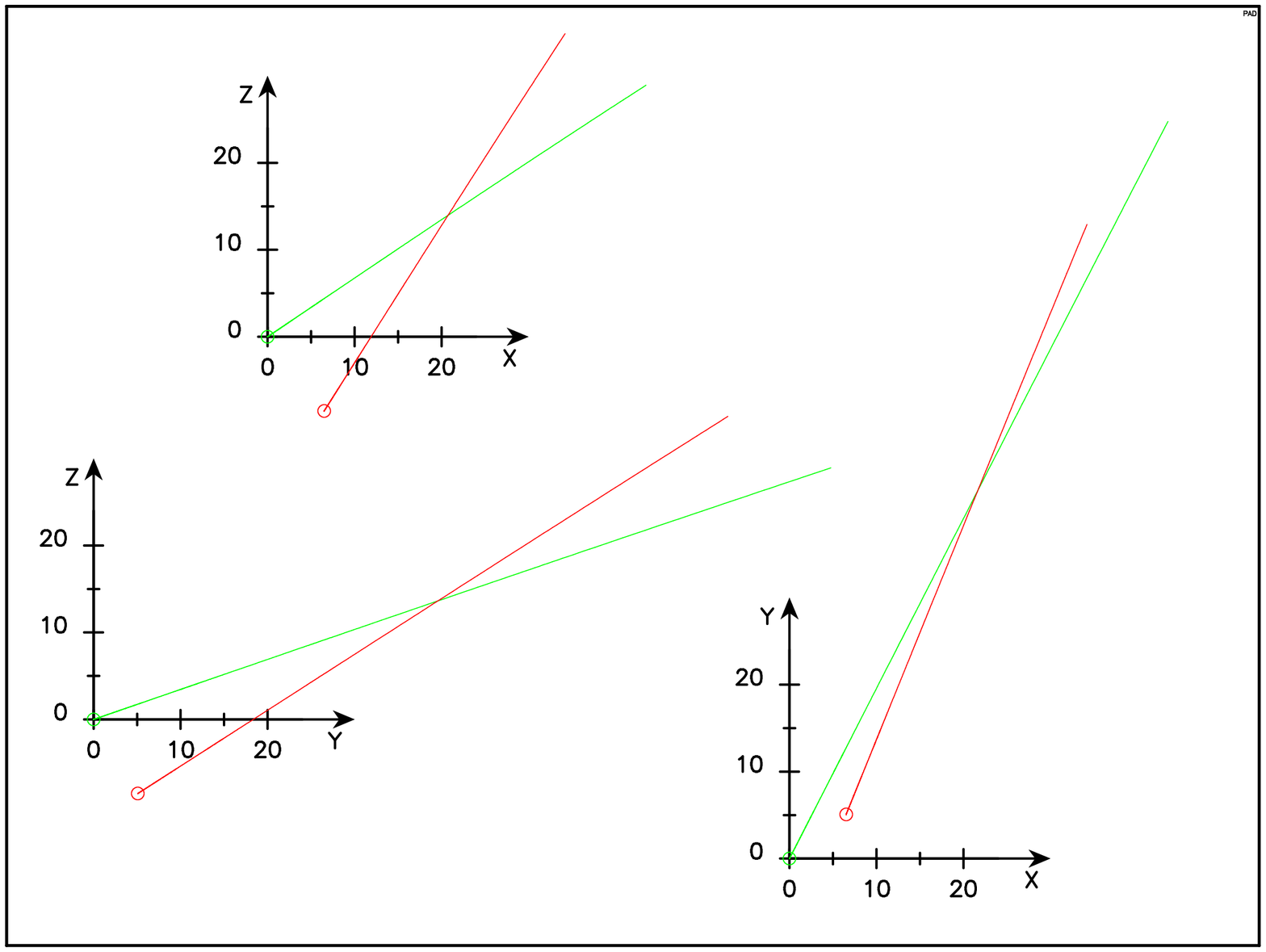}
\label{fig:startrek22hip107556a}
\end{center}
& 
\begin{center}
\includegraphics[angle=270, width=0.42\textwidth]{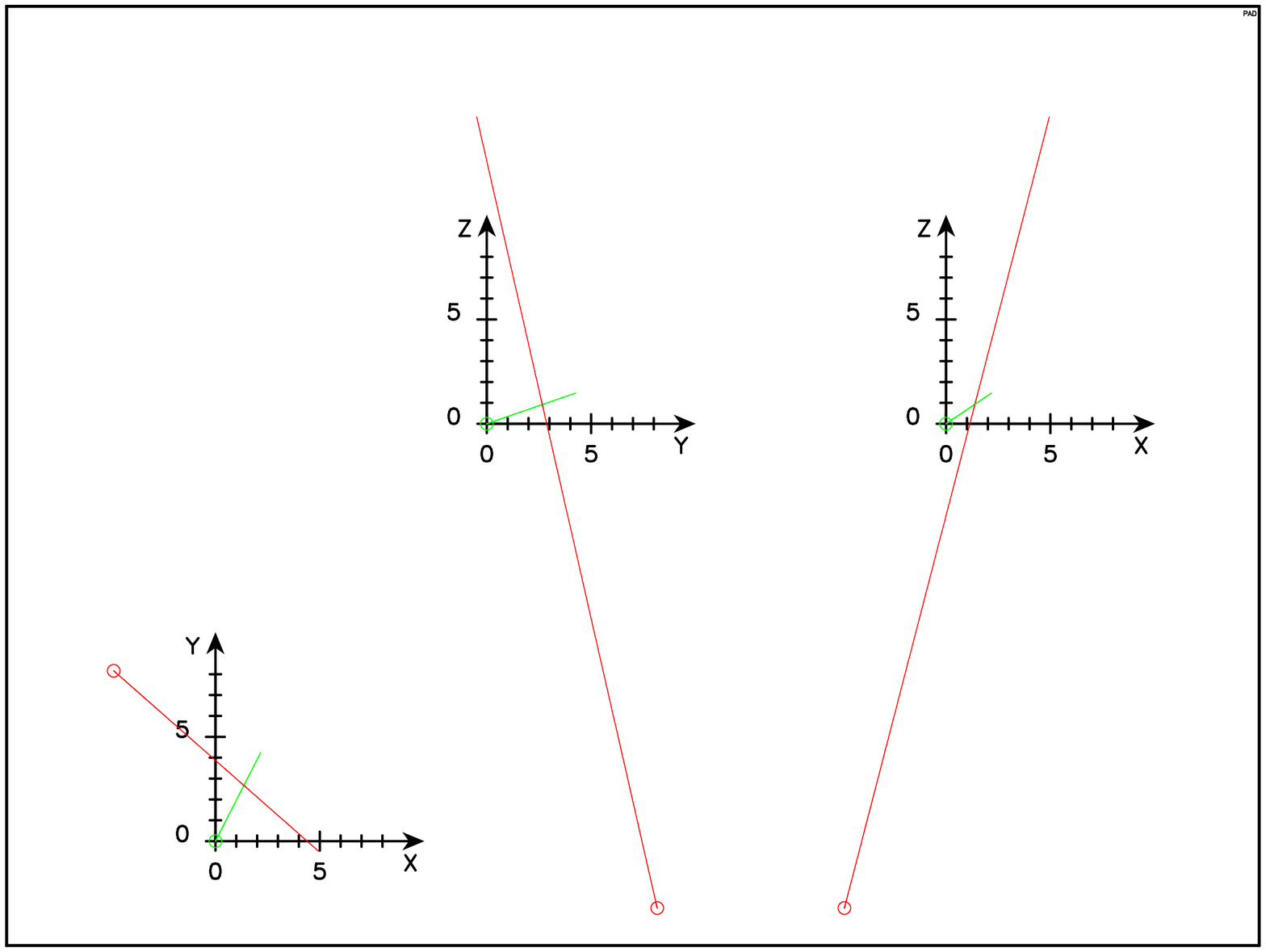}
\label{fig:startrek22hip3757a}
\end{center}
\\

Fig.~A.3. The geometry of the encounter of 'Oumuamua with the star $\delta$\,Capricorni 1.5 Myr ago. Their past motion over 3.74 Myr is shown. 

&	

Fig.~A.4. The geometry of the encounter of 'Oumuamua with the star HIP\,3757 118 kyr ago. 186 kyr of motion of these bodies is shown here.

\\ 
 \begin{center}
\includegraphics[angle=270, width=0.42\textwidth]{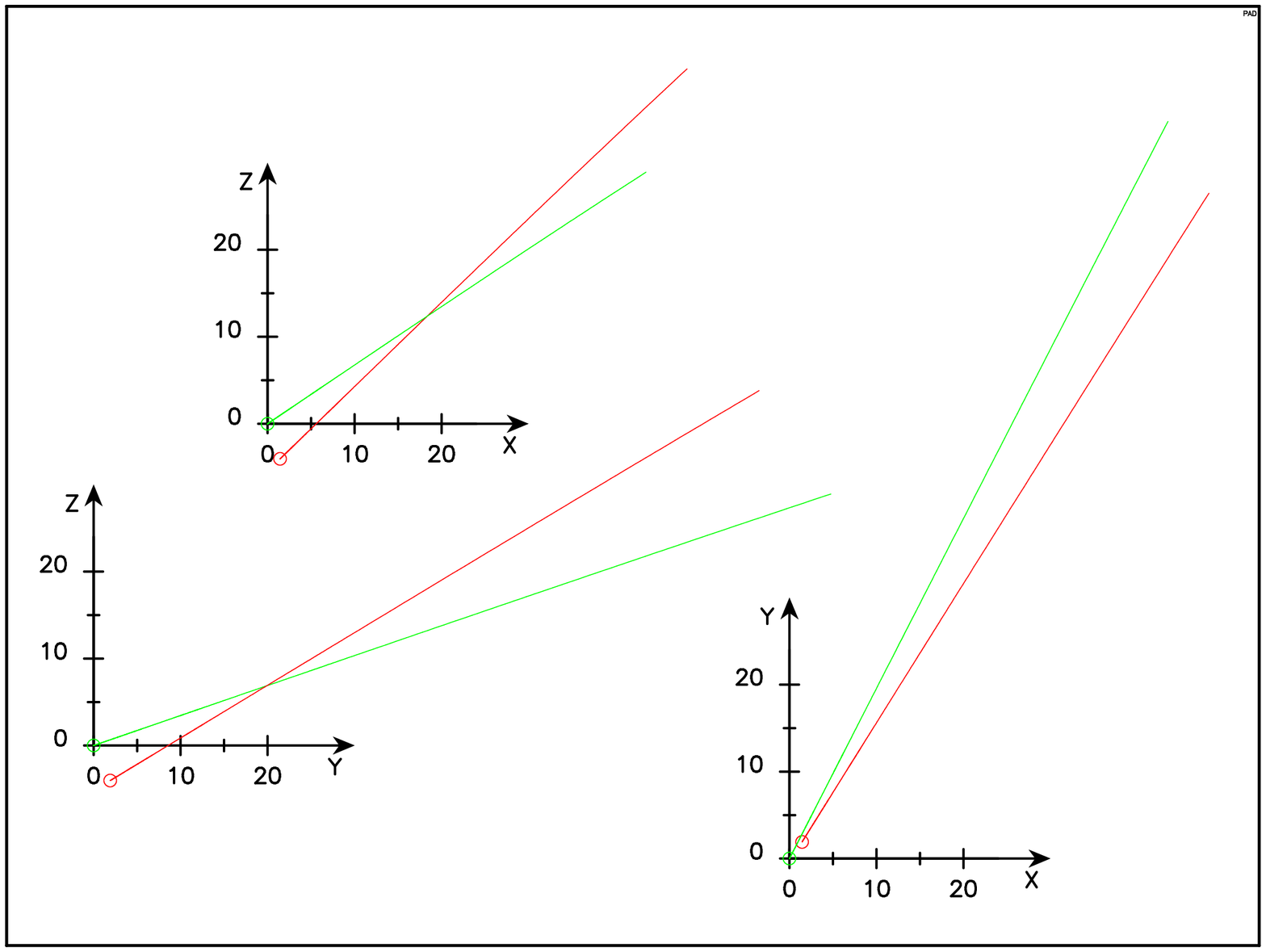}
\label{fig:startrek22hip113020a}
	\end{center}
&
\begin{center}
\includegraphics[angle=270, width=0.42\textwidth]{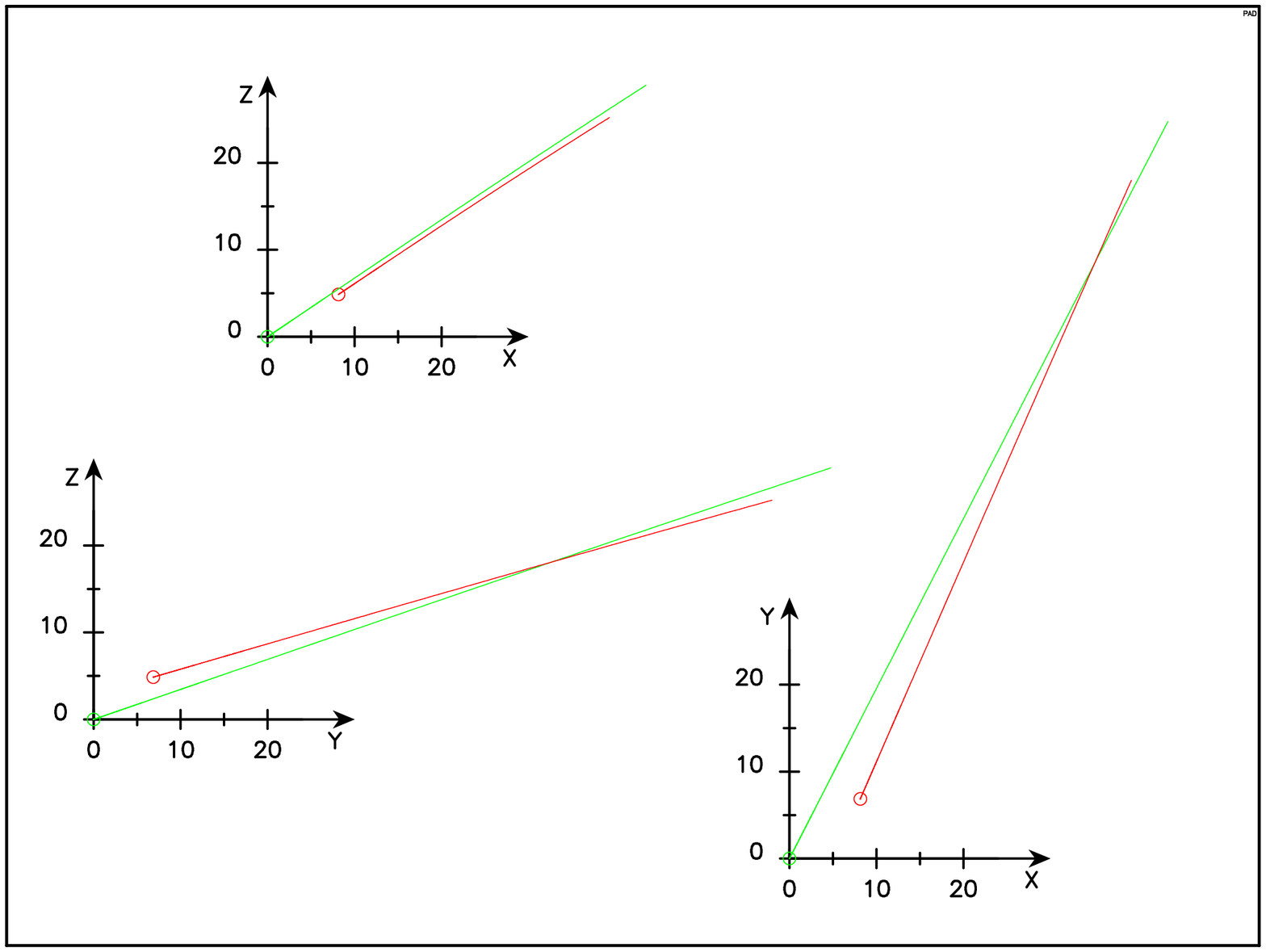}
\label{fig:startrek22ucac4535-065571a}
\end{center}
\\
	
Fig.~A.5. The geometry of the encounter of 'Oumuamua with the star HIP\,113020 0.8 Myr ago. Past motion during 3.74 Myr is shown.
	
&

Fig.~A.6. The geometry of the encounter of 'Oumuamua with the star UCAC4\,535-065571 2.14 Myr ago. Past motion during 3.74 Myr is shown.

\\ 
	
\end{tabular} 

\end{table}

} } 
\end{document}